\documentclass{sig-alternate-05-2015}

\usepackage{times}
\usepackage{paralist}
\usepackage{listings}
\usepackage{etoolbox}
\usepackage{authblk}

\usepackage{fancybox}
\usepackage{graphicx}
\graphicspath{{figures/}}
\DeclareGraphicsExtensions{.pdf,.jpeg,.png}
\usepackage{amsmath}
\usepackage{subfigure}
\usepackage{caption}
\usepackage{fancyvrb}
\usepackage{color}
\usepackage{algorithmic}
\usepackage{array}
\usepackage{mdwmath}
\usepackage{mdwtab}
\usepackage{booktabs}
\usepackage{listings}
\usepackage{eqparbox}
\usepackage{stfloats}
\usepackage{url}
\usepackage{comment}
\usepackage{framed}
\usepackage{ifthen}
\usepackage{balance}
\usepackage{floatrow}

\usepackage[turnoff]{notes}
\usepackage{multirow}
\usepackage{xspace}
\usepackage{listings}
\usepackage[dvipsnames]{xcolor}

\usepackage{blindtext}

\floatstyle{plaintop}
\restylefloat{table}

\newcommand{\rom}[1]{\uppercase\expandafter{\romannumeral #1\relax}}

\newcommand{\etal}{\hbox{\emph{et al.}}\xspace}
\newcommand{\eg}{\hbox{\emph{e.g.,}}\xspace}
\newcommand{\ie}{\hbox{\emph{i.e.}}\xspace}

\newcommand{\wrt}{\hbox{\emph{w.r.t.}}\xspace}

\newcommand{\etc}{\hbox{\emph{etc.}}\xspace}

\newcommand{\tableshrink}[1]{\vspace{-0.25in}}

\definecolor{gray50}{gray}{.5}
\definecolor{gray40}{gray}{.6}
\definecolor{gray30}{gray}{.7}
\definecolor{gray20}{gray}{.8}
\definecolor{gray10}{gray}{.9}
\definecolor{gray05}{gray}{.95}

\newlength\Linewidth
\def\findlength{\setlength\Linewidth\linewidth
\addtolength\Linewidth{-4\fboxrule}
\addtolength\Linewidth{-3\fboxsep}
}
\newenvironment{examplebox}{\par\begingroup
   \setlength{\fboxsep}{5pt}\findlength
   \setbox0=\vbox\bgroup\noindent
   \hsize=0.95\linewidth
   \begin{minipage}{0.95\linewidth}\normalsize}
    {\end{minipage}\egroup
    \textcolor{gray20}{\fboxsep1.5pt\fbox
     {\fboxsep5pt\colorbox{gray05}{\normalcolor\box0}}}
    \endgroup\par\noindent
    \normalcolor\ignorespacesafterend}

\newcounter{RQCounter}
\newcounter{RQACounter}

\newcommand{\RQ}[2]{%
\refstepcounter{RQCounter} \label{#1}
 \begin{center}	
  \begin{examplebox}
   \textbf{RQ\arabic{RQCounter}.}~#2
  \end{examplebox}	 
 \end{center}
}

\newcommand{\RQA}[2]{%
\refstepcounter{RQACounter} \label{#1}
\vspace{0.1in} \noindent\textbf{RQ\arabic{RQACounter}.~#2 \vspace{0.05in}}

}

\newcommand{\RS}[2]{%
\begin{framed}%
\filbreak
\textbf{Result {\ref{#1}}:~}{\emph {#2}}%
\end{framed}
}

\definecolor{javared}{rgb}{0.6,0,0} 
\definecolor{javagreen}{rgb}{0.25,0.5,0.35} 
\definecolor{javapurple}{rgb}{0.5,0,0.35} 
\definecolor{javadocblue}{rgb}{0.25,0.35,0.75} 

\lstdefinestyle{customc}{
  belowcaptionskip=\baselineskip,
  breaklines=true,
  xleftmargin=\parindent,
  language=java,
  showstringspaces=false,
  basicstyle=\scriptsize\ttfamily,
  keywordstyle=\bfseries\color{javapurple},
  commentstyle=\itshape\blue,
  belowskip=-10pt,
  aboveskip=-5pt
}

\newcommand\Red[1]{\textcolor[rgb]{1.00,0.00,0.00}{\textbf{#1}}}

\newcommand\blue[1]{\textcolor[rgb]{0.00,0.00,1.00}{{#1}}}

\newcommand\dkgreen[1]{\textcolor[rgb]{0.0,0.6,0}{\textbf{#1}}}

\newcommand{\ngm}{\$gram\xspace}
\newcommand{\ngmt}{\$gram+type\xspace}
\newcommand{\sbfl}{\hbox{${\cal S}\hspace{-0.01in}{\cal B}\hspace{-0.01in}{\cal B}\hspace{-0.01in}{\cal L}$}\xspace}
\newcommand{\LM}{\hbox{${\cal L}\hspace{-0.01in}{\cal M}$}\xspace}
\newcommand{\tool}{EnSpec\xspace}

\newcommand{\bug}{bug\xspace}

\newcommand{\Bug}{Bug\xspace}
\newcommand{\buggy}{buggy\xspace}

\newcommand{\passing}{passing\xspace}
\newcommand{\failing}{failing\xspace}

\newcommand{\defj}{Defects4J\xspace}
\newcommand{\mbug}{ManyBugs\xspace}

\newcommand{\chart}{JFreechart\xspace}
\newcommand{\closure}{Closure compiler\xspace}
\newcommand{\lang}{Apache commons-lang\xspace}
\newcommand{\maths}{Apache commons-math\xspace}
\newcommand{\jtime}{Joda-Time\xspace}

\newcommand{\Comment}[1]{}

\newcommand{\ripon}[1]{{\scriptsize \todo{Ripon:  {\color{brown} #1}}}}

\newcommand{\aucecen}{{\it {\bf $AUCEC_{en}$}}\space}
\newcommand{\aucecsp}{{\it {\bf $AUCEC_{sp}$}}\space}

\begin{document}



\title{Entropy Guided Spectrum Based Bug Localization Using Statistical Language Model}

\author[1,*]{Saikat Chakraborty}
\author[2,*]{Yujian Li}
\author[3,*]{Matt Irvine}
\author[+]{Ripon Saha}
\author[5,*]{Baishakhi Ray}
\affil[*]{Department of Computer Science, University of Virginia, Charlottesville, VA - 22903}
\affil[1,2,3,5]{\{saikatc, yl7kd, mji7wb, rayb\}@virginia.edu }
\affil[+]{Fujitsu Laboratories of America, Sunnyvale, CA - 94085}
\affil[+]{rsaha@us.fujitsu.com}

\maketitle

\begin{abstract}
Locating bugs is challenging but one of the most important activities in software development and maintenance phase because there are no certain rules to identify all types of bugs. Existing automatic bug localization tools use various heuristics based on test coverage, pre-determined buggy patterns, or textual similarity with bug report, to rank suspicious program elements. However, since these techniques rely on information from single source, they often suffer when the respective source information is inadequate.  For instance, the popular spectrum based bug localization may not work well under poorly written test suite. In this paper, we propose a new approach, \tool, that guides spectrum based bug localization using code entropy, a metric that basically represents {\it naturalness} of code derived from a statistical language model. Our intuition is that since buggy code are high entropic, spectrum based bug localization with code entropy would be more robust in discriminating buggy lines vs.~non-buggy lines. We realize our idea in a prototype, and performed an extensive evaluation on two popular publicly available benchmarks. Our results demonstrate that \tool outperforms a state-of-the-art spectrum based bug localization technique.   
\end{abstract}






\section{Introduction}
\label{sec:intro}

Localizing bugs is an important, time consuming, and expensive process, especially for a system at-scale. Automatic bug localization can play an important role in saving developers' time in debugging, and thus, may help developers fixing more bugs in a limited time. Using various statistical and program analysis approaches, these bug localization techniques automatically identify suspicious code elements that are highly likely to contain bugs. Developers then manually examine these suspicious code to pinpoint the bugs. 

Existing bug localization techniques can be broadly classified into two categories: i) test coverage-based dynamic approaches~\cite{jones2005tarantula,abreu2009practical,zeller2002simplifying,cleve2005locating,liblit2005scalable,liu2005sober}, and ii) pattern-based~\cite{copeland2005pmd,findbugs, Engler:2001:SOSP,Chelf:2002:Paste} or information retrieval-based (IR) static approaches~\cite{rao2011retrieval,zhou2012should,saha2013improving,ye2014learning}. Dynamic approaches first run all the test cases, and then analyze the program statements covered by \passing and \failing test cases. For example, spectrum based bug localization (\sbfl), a popular dynamic bug localization technique, prioritizes the program elements for debugging that are executed more by \failing test cases than by \passing test cases. In contrast, static approaches do not run any test cases. Rather, it searches for some previously known buggy patterns in source code or looks for buggy files based on bug reports. 

Both of these bug localization approaches have their own set of advantages and disadvantages. For instance, static methods are often imprecise or inaccurate. On the other hand, the accuracy of dynamic approaches is highly dependent on the quality (code coverage, \etc) of the test suite. In real world projects, most of the test suite may not have enough code coverage to locate bugs efficiently. Therefore, in many cases, developers do not get the full benefit of bug localization techniques~\cite{johnson2013don} and have to significantly rely on manual effort and prior experiences. 


Besides static and dynamic properties of a program, it has also been observed that how developers write code is also important for code quality~\cite{hellendoorn2015will}.  Real-world software that are developed by regular programmers tend to be highly repetitive and predictable~\cite{gabel2010study}. Hindle \etal was the first to show that such repetitiveness can be successfully captured by a statistical language model~\cite{hindle2012naturalness}. They called this property as {\it naturalness} of code and measured it by standard information theory metric {\it entropy}. The less entropy a code snippet exhibits, the more the code is natural. Inspired by this phenomena, Ray \etal~\cite{ray2016naturalness} investigated if there is any correlation between buggy code and entropy. They observed that buggy codes are in general less natural, \ie they have higher entropy than non-buggy code. 

In this paper, our key intuition is that, since the high entropic code tends to be buggy~\cite{ray2016naturalness,campbell2014syntax,wang2016bugram}, code entropy can be an effective orthogonal source of information to \sbfl to improve the overall accuracy of bug localization. This notion seems to be plausible since from a set of suspicious code, as reported by a  standard \sbfl technique, experienced programmers often intuitively identify the actual bugs because buggy code elements are usually a bit unnatural than rest of the corpus. If entropy can improve \sbfl, it would be particularly useful when a test suite is not strong enough to discriminate buggy lines or when the suspicious scores of many lines are the same. Furthermore, to realize this hybrid approach, we only need source code (no other external meta-source), which is always available to the developers. 

To this end, we introduce \tool, that automatically calculates the entropy for each program line and combines with a state-of-the-art \sbfl using a machine learning technique to return a ranked list of suspicious lines for investigation.
Here, we studied bug localization at line granularity to ensure maximum benefit to the developers, although locating bugs in method and file-granularity is also possible. We performed an extensive evaluation of \tool on two popular publicly available bug-dataset: \defj~\cite{just2014defects4j} and \mbug~\cite{le2015manybugs} written in Java and C respectively. In total, we studied more than 500 bugs (3,715 buggy lines) bugs, around 4M LOC from 10 projects.

Overall, our findings corroborate our hypothesis that entropy can indeed improve bug localization capability of spectrum based bug localization technique. We further evaluate \tool for both C and Java projects showing that the tool is not programming language dependent. In particular, our results show that:

\begin{itemize}
\item Entropy score, as captured by statistical language models, can significantly improve bug localization capability of standard \sbfl technique. 

\item Entropy score also boosts \sbfl in cross-project bug localization settings.
\end{itemize} 

In summary, we make the following contributions in this paper:

\begin{enumerate}
\item We introduce the notion of entropy in spectrum based bug localization.

\item We present \tool that effectively combine entropy score with the suspicious score of spectrum based bug localization using a machine learning technique.

\item We provide an extensive evaluation of \tool on two publicly available benchmarks that demonstrates the effectiveness of our approach.
\end{enumerate}


\section{Preliminaries}
\label{sec:prelim}
In this section, we discuss the preliminaries and backgrounds of our work. 

\subsection{Spectrum-based \Bug Localization}
\label{subsec:spectrum}

Given a \buggy code-base with at least one \bug reproducing test case, a spectrum-based \bug localization technique (\sbfl) ranks the code elements under investigation (e.g., files/classes, functions/methods, blocks, or statements) based on the execution  traces of passing and failing test cases. Therefore, in this approach, first the subject program is instrumented at an appropriate granularity to collect the execution trace of each test case ({\it test spectra}). The basic intuition behind \sbfl is that the more a code element appears in failing traces (but not in passing traces), the more suspicious the element to be \buggy.




\begin{table}[!htbp]
  \centering
  \scriptsize
  \caption{{\textbf{\small Spectrum based suspiciousness score of program element e}}} 
    \begin{tabular}{llp{1.2cm}l}
    \toprule
      & \textbf{\#passed} & \textbf{\#failed} & \textbf{spectrum} \\
      &     \textbf{test} &  \textbf{test} &  \textbf{score} \\
    \midrule
 \textbf{total tests} & $P$ & $F$  & {$S=Func(e_p,e_f,n_p,n_f)$}  \\
    &&& \\
\textbf{e executed}  &$e_p$& $e_f$& 
{$S_{Tarantula} = \frac{\frac{e_f}{F}}{\frac{e_f}{F}+\frac{e_p}{P}}$}        \\
  &&& \\
\textbf{e not executed} & $n_p$=$P$-$e_p$ & $n_f$=$F$-$e_f$ &    
$S_{ochai} = \frac{e_f}{\sqrt{(e_f+e_p)(e_f+n_f)}}$ \\
       &&& \\
    \bottomrule
    \end{tabular}%
  \label{tab:spectra}%
\end{table}%

More specifically, for a given program element $e$, \sbfl records how many test cases execute and do not execute $e$, and computes the following four metrics: the number of (i) tests passed ($e_p$) and (ii) tests failed ($e_f$) that executed $e$, and the number of (iii) tests passed ($n_p$) and (iv) tests failed ($n_f$) that did not execute $e$. 
A suspiciousness score is calculated is calculated as a function of these four metric: $S = Func(e_p,e_f,n_p,n_f)$, as shown in Table~\ref{tab:spectra}.
The table also presents two widely used suspiciousness score measure: Tarantula and Ochai. \sbfl ranks the program elements in a decreasing order of suspiciousness and presents to the developers for further investigation to fix the bug~\cite{xuan2014learning}. These scores also help to repair program automatically~\cite{le2012representations}.


\subsection{Language Models}

Although real-world software programs are often large and complex, the program constructs (such as tokens) are repetitive, and thus provide useful predictable statistical property~\cite{Hindle:2012:ICSE, Raychev:2014:PLDI, Tu:2014:FSE, Franks:2015:ICSE}. These statistical properties of code resemble natural languages, and thus, natural language models can be leveraged for software engineering tasks. 


\textbf{Cache based N-gram Model (\ngm):} Hindle \etal introduced n-gram model for software code~\cite{hindle2012naturalness}, which is essentially an extension of n-gram language model used in natural language processing tasks based on the \textit{Markov independence assumption}~\cite{brown1992class_ngram}.  If a sequence, $s$ consists of $m$ tokens ($a_1 a_2 a_3 ... a_m$), according to the \textit{Full Markov Model}, the probability, $p(s)$, of that sequence is given in Equation~\ref{eqMkvModel}
\begin{equation}
\label{eqMkvModel}
\small
p(s) = p(a_1)p(a_2|a_1)p(a_3|a_1a_2) ... p(a_m|a_1a_2...a_{m-1})
\end{equation}

\textit{N-gram model} is a simplification of full Markov model based on the assumption that every token is dependent on previous $n-1$ token, where $n$ is a model parameter. Essentially with $n=\infty$, the model converges to the full Markov model. Since, actual probabilities are very difficult to find, researchers often use empirical probabilities to represent actual probabilities, which is highly dependent on the training data. Initially, the probability of a token or any n-gram which is not seen in the corpus will be \textit{zero} resulting the total probability to be \textit{zero}. To overcome this problem Hindle \etal~\cite{hindle2012naturalness} also adopted smoothing techniques from natural language processing literature.

Tu \etal~\cite{tu2014localness} further improved the above model based on the observation that source code tends to be highly localized, \ie  particular token sequences may occur often within a single file or within particular classes or functions. They proposed a \ngm model that introduces an additional cache\textemdash list of n-grams curated from local context and used them in addition to a global n-gram model. They also defined the entropy of a code sequence $S$ by language model $M$ by Equation~\ref{eqLocalModel}:

\begin{equation}
\label{eqLocalModel}
\small
H_M(S) = -\frac{1}{N}\log_2p_M(S) = -\frac{1}{N}\sum\limits_{i=1}^{N}\log_2P(t_i|h) 
\end{equation}

\textbf{Language model to predict buggy code:} Ray \etal~\cite{ray2016naturalness} demonstrated that there is a strong negative correlation between code being buggy and the naturalness of code. When a simple \ngm model is trained on previous versions of project source code and applied to calculate naturalness of code snippet, buggy codes are shown to exhibit higher unnaturalness than the non-buggy codes. They introduced a syntax-sensitive entropy model to measure naturalness of code. In their investigation, they found that some token types such as packages, methods, variable names are less frequent, and hence high entropic than others. They normalized the entropy score and derived a Z-score with line-type information from the program's Abstract Syntax Tree (AST). The Z-score is defined as:

\begin{equation}
\label{eqZscore}
\small
\ngmt = \frac{entropy_{line}-\mu_{type}}{SD_{type}}
\end{equation}

In Equation~\ref{eqZscore}, $\mu_{type}$ is the mean \ngm model entropy of a given line type, and $SD_{type}$ is the standard deviation of that line type. This Z-score gives the syntax-sensitive entropy. They reported that, buggy lines of codes are usually unnatural and highly entropic. With further investigation, they also found that, when developers fixed those buggy lines of codes, the entropy of the code decreased. In this work, we used state of the art \ngm model along with syntax-sensitive entropy model.

\section{Motivating Example}
\label{sec:motivation}
In this section, we present a real-world example that motivated us to incorporate the \Comment{Natural Language Property} {\it naturalness} property of code  (i.e. {\it entropy} based features) in \sbfl to overcome a key limitation of \sbfl. 

The main limitation of testing based \bug localization approaches, such as \sbfl, is that the quality of their results highly depend on the quality of test cases. If the passing test cases have low code coverage, an \sbfl tool may return a large number of program elements with high suspiciousness score, most of which are false positives. However, generating an adequate test suite is incredibly difficult. Therefore, in many cases \sbfl performs poorly. 

\begin{table}[h]
\begin{tabular}{p{0.9\textwidth}}

%
\scriptsize {\tt{--- /Closure/89/buggy/src/com/google/javascript/jscomp/}}
\scriptsize {\tt{GlobalNamespace.java}}\\
\scriptsize {\tt{+++ /Closure/89/fix/src/com/google/javascript/jscomp/}}
\scriptsize {\tt{GlobalNamespace.java}}
\begin{lstlisting}[ language=java]

- if (@\Red{type != Type.FUNCTION \&\& aliasingGets > 0}@) { 
	//Entropy before patch 7.36
+ if (@\dkgreen{aliasingGets > 0}@) { 
	//Entropy after patch 1.15
	return false;
}
\end{lstlisting}\\
\end{tabular}
\caption{Effectiveness of entropy based features to improve \sbfl}
\label{tbl:motivating_example}
\end{table}
\vspace{20pt}


\Comment{For example, on a test run, a trained model is tested with 3731 lines of which 10 lines are actual bug. Test result shows that 9 of those 10 buggy lines are placed relatively higher position in the ranklist when we incorporated {\it Entropy} based features.} 

Table~\ref{tbl:motivating_example} presents a patch that fixed a bug in \closure (Defects4J \bug ID: 89). The buggy line, marked in~\Red{red}, was never used in the existing corpus before. Hence, the line was unnatural to a \LM with a high entropy score of $7.36$. When developer fixed the bug (see~\dkgreen{green} line), the code becomes more natural with a reduced entropy score of $1.15$. A traditional state-of-the art \sbfl technique 
placed the buggy line at 57$^{th}$ position, while \tool using both {\it entropy} and {\it spectrum} based features, placed the line at 12$^{th}$ position in the ranked list of suspicious lines. This shows that entropy of code, as derived from \LM, can play an important role to improve the ranking of the actual \buggy lines.

\section{Proposed Approach}
\label{sec:method}

\begin{figure*}[!htpb]
\centering
\includegraphics[width=\textwidth]{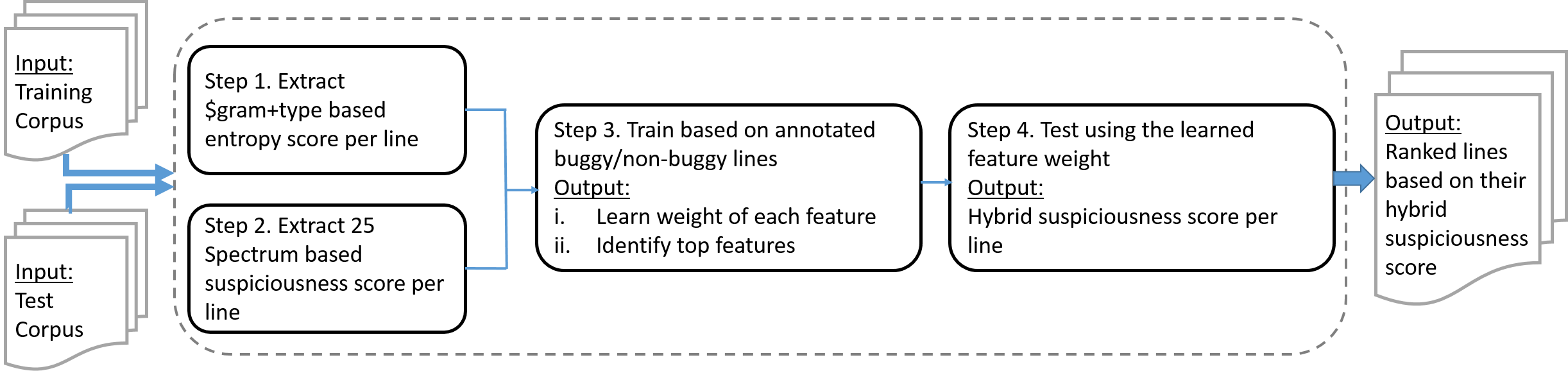}
\caption{\textbf{\small \tool Workflow}}
\label{fig:schema}
\end{figure*}
\vspace{15pt}

In this section, we describe our tool, \tool. An overview of our approach is shown in Figure~\ref{fig:schema}. The goal of \tool is to localize bugs using a hybrid bug localization technique: a combination of dynamic spectrum based bug localization (\sbfl) and static natural language model based defect prediction (\LM).  \tool takes two sets of code corpus as input\textemdash training and testing set. Next, \tool works in following four steps: 
{\em Step-1.} \tool collects entropy score per code element based on a language model for each input project.
{\em Step-2.} For each project version in the training and test corpus,  \tool records test coverage and collects various \sbfl based suspiciousness scores per code element. 
{\em Step-3.}  In this step, \tool learns from the training data, how the suspiciousness scores and entropy collected in above two steps relate to buggy/non-buggy classes and learns feature weight. In Section~\ref{subsec:dc}, we describe the data collection phase in more detail: how we annotate each code element as buggy/non-buggy.
{\em Step-4.} Based on the learned feature-weight, \tool assigns a suspiciousness score of each code element in the test corpus. The suspiciousness score depicts the probability of a code element to be buggy. 
Finally, the output of \tool is a ranked list of code elements based on their decreasing suspiciousness score. 

In theory, \tool should work on code elements at any granularity\textemdash line, method, file, etc. In this paper, we use \tool to localize bugs at a line granularity. In the following section, we describe these steps in details.

\subsection*{Step-1:~Generating entropy using \LM}

For generating entropy per program line, we adopted the \ngm language model proposed by Tu \etal~\cite{tu2014localness}. For every line in source code we calculated following three entropy values: 

1.~{Forward Entropy $(E_f)$}: Entropy value of a token is calculated based on the probability of seeing the token given its prefix token sequences.  We calculate this entropy by parsing the file from beginning to end, \ie considering the token sequences as it is in the source file.

2.~{Backward Entropy $(E_b)$}:  Entropy value of a token is calculated based on the probability of seeing the token given its suffix token sequences. We calculate this entropy by parsing the file in reverse order, \ie from end to beginning. 

3.~{Average Entropy $(E_a)$}: This entropy value is calculated as the average of $E_f$ and $E_b$.

We use these three values as our \LM based features. We further normalized these values based on their AST type, as shown in Equation~\ref{eqZscore}. We refer these three normalized entropy values as \textit{entropy related features} in the rest of the paper.


\begin{table*}[!t]
\centering
\scriptsize
\caption{{\textbf{\small 25 \textit{suspiciousness} scores from literature that are used as \sbfl features in \tool}}}
\label{tbl:rank_metrics}
\begin{tabular}{p{0.9cm}c|p{0.9cm}c|p{0.9cm}c|p{1.1cm}c}
\toprule
Metric & Formula & Metric & Formula & Metric & Formula & Metric & Formula\\
\midrule
Tarantula & $\frac{\frac{e_f}{e_f+n_f}}{\frac{e_f}{e_f+n_f} + \frac{e_p}{e_p+n_p}}$ & Ochiai & $\frac{e_f}{\sqrt{(e_f+e_p)(e_f+n_f)}}$ & Jaccard & $\frac{e_f}{e_f + e_p + n_f}$ & SimpleMatching & $\frac{e_f+n_p}{e_f + e_p + n_f+n_p}$\\
\midrule
 S$\phi$rcenDice & $\frac{2e_f}{2e_f + e_p + n_f}$ & Kulczynskil & $\frac{e_f}{e_p + n_f}$ & RusselRao & $\frac{e_f}{e_f + e_p + n_f + n_p}$ & RogersTanimoto & $\frac{e_f+n_p}{e_f + e_p + 2n_f + e_P}$ \\
 \midrule
 M1 & $\frac{e_f+n_p}{e_p + n_f}$ & M2 & $\frac{e_f}{e_f+n_p+2e_p+2e_f}$  & Overlap & $\frac{e_f}{min(e_f, e_p, n_f)}$ & Ochiai2& $\frac{e_fn_p}{\sqrt{(e_f+e_p)(n_f+n_p)(e_f+n_p)(e_p+n_f)}}$\\
 \midrule
 Dice & $\frac{2e_f}{e_f+e_p+n_f}$ & Ample & $\left| \frac{e_f}{e_f+n_f} - \frac{e_p}{e_p+n_p}\right|$ & Hamann & $\frac{e_f + n_p - e_p - n_f}{e_f + e_p + n_f + n_p}$ &  Zoltar & $\frac{e_f}{e_f+e_p+n_f+\frac{10000n_fe_p}{e_f}}$\\
 \midrule
 Goodman & $\frac{2e_f-n_f-e_p}{2e_f+n_f+e_p}$ & Sokal & $\frac{2e_f+2e_p}{2e_f+2e_p+n_f+n_p}$ & Hamming & $e_f + n_p$ & Kulczynski2 & $\frac{1}{2}\left(\frac{e_f}{e_f+n_f} + \frac{e_f}{e_f+n_p}\right)$\\
 \midrule
 Euclid & $\sqrt{e_f+n_p}$ & Anderberg & $\frac{e_f}{e_f+2e_p+2n_f}$ & 
 Wong1 & $e_f$ & Wong2& $e_f - e_p$\\
 \midrule
Wong3 & \multicolumn{6}{l}{
$e_f -h,~where~h = \left\{
\begin{array}{ll}
e_p & \text{if}~e_p\leq 2\\
2+0.1(e_p-2) & \text{if}~2 \leq e_p \leq 10\\
2.8 +0.01(e_p-10) & \text{if}~e_p\geq 10\\
\end{array}\right.
$ 
} & \\
\bottomrule

\end{tabular}
\end{table*}

\subsection*{Step-2:~Extracting suspiciousness score using \sbfl techniques.}
For all the input project versions, we first instrument the source code to record program execution traces, or coverage data. Both \defj and \mbug dataset provide APIs for collecting such coverage data. Then, to collect the execution traces, we extract the test classes and test methods from the project source code and run the test cases. We record the execution traces for each test case with its passing/failing status. These test spectra characterize the program's behavior across executions by summarizing how frequently each source code line was executed for passing and
failing tests. Now for each line we  calculate 4 values $(e_p, e_f, n_p, n_f)$, as described in Section~\ref{sec:prelim}. Next, using these 4 values, we generate 25 suspiciousness scores, as  described by Xuan \etal~\cite{xuan2014learning}. We use these 25 scores(see Table~\ref{tbl:rank_metrics}) as our \sbfl features. 

The next two steps implement the training and testing phase of a classifier based on buggy and non-buggy program lines. We adapted Li~\etal's learning to rank algorithm for this purpose~\cite{li2007mcrank}.

\subsection*{Step-3:~Training Phase}
 Given a set of buggy and non-buggy lines, \tool learns the relation  between \sbfl and entropy related features on the bugginess of program lines.

First, all lines in the training dataset were annotated with a relevance score of bugginess: $R_b$ for each buggy line, and $R_g$ for each non-buggy line, where $R_b > R_g$.  Thus, each line $l$ in the training code corpus is represented as a tuple, $\langle Sp_l, En_l, R_l \rangle$, where $Sp_l$ is a set of \sbfl features, $En_l$ is a set of entropy related features, and $R_l \in \{R_b, R_g\}$ is the bug-relevance score. 
Then, we pass the whole corpus to a  machine learner, which learns the probability distribution $P(R_l|Sp_l,En_l)$ of the relevance scores given the feature values.

\subsection*{Step-4:~Testing Phase}

In testing phase, for each line $l$ in the test corpus, we compute a suspiciousness score ($susp$) based on equation ~\ref{eqn:proba_suspiciousness_score}.

\begin{equation}
\label{eqn:proba_suspiciousness_score}
\begin{split}
susp_{l} = f(R_b) &* P(R_b|Sp_{l},En_{l})\\
&+ f(R_g) * P(R_b|Sp_{l},En_{l})
\end{split}
\end{equation}

Here, $f(R_b)$ and $f(R_g)$ are monotonically increasing functions. To keep things simple, we used identity function, \ie, $f(R_b) = R_b$ and $f(R_g) = R_g$, which is monotonic as well. This transforms equation ~\ref{eqn:proba_suspiciousness_score} into: 

\begin{equation}
\label{eqn:expected_value_suspiciousness}
\begin{split}
susp_{l} &= \sum_{R_l \in \{R_b, R_g\}}{R_l * P(R_l|Sp_{l},En_{l})} \\
&= \mathbb{E}[R_l|Sp_{l}, En_{l}] 
\end{split}
\end{equation}

We use ensemble~\cite{dietterich2002ensemble} of $M$ different models trained on randomly sampled subset of original dataset. Each model $M_k$ computes a suspiciousness score $susp_{l}^{k}$,  based on the expected relevance score of equation~\ref{eqn:expected_value_suspiciousness}. Our final hybrid suspiciousness score is calculated by equation~\ref{eqn:ensemble}:

\begin{equation}
\label{eqn:ensemble}
HySusp_{l} = \frac{1}{M}\sum_{k=1}^{M}{susp_{l}^k}
\end{equation}

\tool outputs a ranked list of source code lines based on the decreasing order of hybrid suspiciousness score (HySusp), line with highest suspiciousness tops the list. 

\section{Experimental Setup}
\label{sec:experiment}

In this section, we describe how we setup our experiment to evaluate \tool. In particular, we describe the subject systems, how we collect data, evaluation metric, and research questions to evaluate \tool.


\begin{center}
\begin{table*}[!htpb]
\caption{\textbf{\small Study Subjects}}
\label{tab:ssubj}
\centering
\scriptsize
\begin{tabular}{l|l|c|c|c|c|c}
Dataset 			& \textbf{Project} & \textbf{KLoc}	& \textbf{\#Tests}  &	\textbf{\#Bugs} & \textbf{\#Buggy Lines} & \textbf{\#Buggy Lines}\\  
 			         & 	     &		&  &	 & (original dataset) & (original + evolutionary)\\  \toprule
        			& \chart 	     &	96		&	2,205 &	26	& 22 & 32 \\ 
        			& \closure	     &	90		&	7,927 &	133	& 64 & 586\\
\textbf{Defects4J}~\cite{just2014defects4j}  & \maths &	85	&	3,602 &	106	& 38 & 816 \\
(Language: Java)	& \jtime 	     & 28  &	4,130  &	27	 & 46 & 97 \\
                	& \lang			 &	22 &   2,245  &	65	& 50 & 230\\ \midrule
					& \textbf{Total} &	321		&	20,109 & 357 & 220 & 1761\\
\toprule
					& Libtiff		& 77	& 78 &	24	& 944 & -\\
					& Lighttpd		& 62	& 295 &	9	& 26 & -\\
\textbf{ManyBugs}~\cite{le2015manybugs}   & Php		& 1,099	& 8,471 &	104 & 503 & -\\
(Language: C)    	& Python		& 407	& 355 &	15	& 93 & -\\
					& Wireshark		& 2,814	& 63 &	8	& 388 & -\\ \midrule
					& \textbf{Total}	& 4,459	& 9,262 &	160	& 1954 & -\\
\bottomrule

\end{tabular}
\end{table*}
\end{center}

\subsection{Study Subject}
\label{subsec:study_subject}

We used two publicly available bug dataset: \defj~\cite{just2014defects4j} and 
\mbug~\cite{le2015manybugs} (see Table~\ref{tab:ssubj}).  \defj dataset contains $5$ open source projects with $321$K lines of code, $357$ reproducible bugs, and $20$K total number of tests. All the \defj projects are written in Java. We also studied $5$ projects from \mbug benchmark dataset~\cite{le2015manybugs}. These are medium to large open source C projects, with total $4459$K lines of code, $160$ reproducible bugs, and $9262$ total test cases. In both datasets, each bug is associated with a buggy and its corresponding fix program versions. There are some failing test cases that reproduce the bugs in the buggy versions, while after the fixes all the test cases pass. The dataset also provides APIs for instrumentation and recording program execution traces.

\subsection{Data Collection}
\label{subsec:dc}

Here we describe how we identified the buggy program statements. We followed two techniques as described below:

\rom{1}.~\textit{Buggy lines retrieved from original dataset.} We compared each buggy program version with its corresponding fixed version; the lines that are deleted or modified in the buggy version are annotated as buggy program statements. To 
get the differences between two program versions, we used \defj APIs for \defj dataset. For every snapshot, \mbug dataset provides {\tt diff} of different files changed during the fix revision. We directly used those {\tt diff} files. 

We notice that some bugs are caused due to the error of omission\textemdash tests fail due to missing features/functionalities in the buggy version. Fixing these bugs do not require deleting or modifying program statements in the buggy version, but adding new lines in the fixed version. In such cases, we cannot trivially annotate any of the existing lines in the buggy version as `buggy.' We filter out such bugs from further consideration, as our goal in this work to locate existing buggy lines, as opposed to detecting error of omission. Table~\ref{tab:ssubj} shows the total number of buggy lines per project in the original dataset. 

\begin{figure}[!htpb]
  \centering
  \includegraphics[width=0.9\columnwidth]{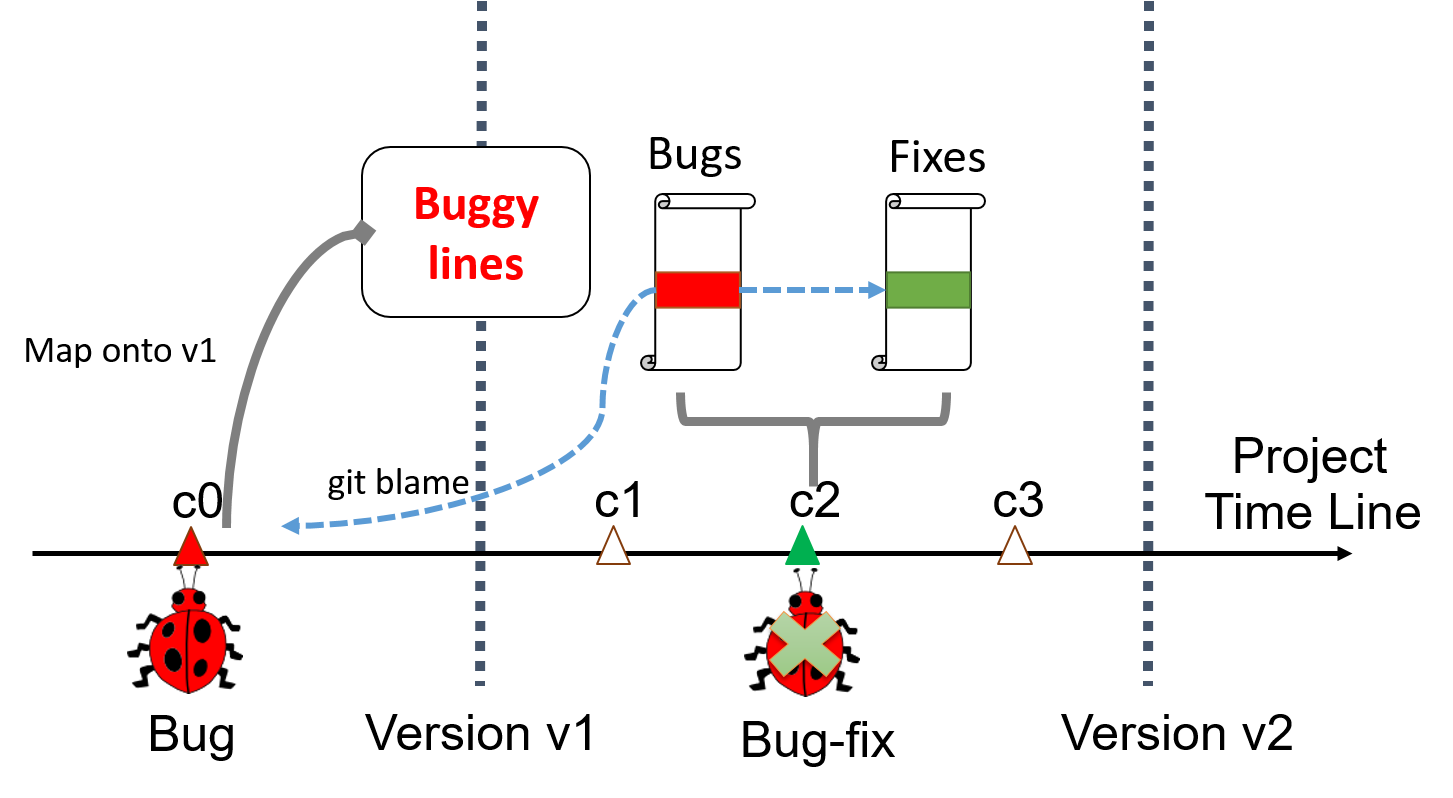}
  \caption{{\small{\bf Evolutionary bug data retrieval: vertical dashed lines correspond to buggy project versions under investigation, each triangle represents project commit (c0$\ldots$c3). For every bug-fix commit (\eg c2), as identified by keyword search,  we first git-blame the buggy lines to identify the original bug-introducing commit (\eg c0) and then map them to the corresponding project versions.}(Adopted from Ray \etal~\cite{ray2016naturalness})}}
  \label{fig:evolution}
\end{figure}

\rom{2}.~\textit{Buggy lines retrieved from project evolution.}
As shown in Table~\ref{tab:ssubj}, the percentage of buggy lines \wrt the total lines of code in \defj dataset is very small (0.07\%). Such unbalanced data of  buggy vs.~non-buggy lines poses a threat to the efficiency of any classification probelm~\cite{japkowicz2002class,chawla2004editorial}. To reduce such imbalance and thus increase the effectiveness of \sbfl, previous work injects artificial bugs in the software system under tests~\cite{gunneflo1989evaluation,segall1995fiat}. However, since the motivation of this research comes from the findings that bugs are {\em unnatural}~\cite{ray2016naturalness}, artificially introducing bugs like our predecessors may question our conclusions. To overcome this problem, we injected bugs that developers introduced in the source code in reality\textemdash we collected such bugs from project evolutionary history. 

We adopted the similar strategy described in Ray \etal~\cite{ray2016naturalness}. First, we identified bug-fix commits by searching a project's commit logs using bug-fix related keywords: `error', `bug', `fix', `issue', `mistake', `flaw', and `defect', following the methodology described by Mockus~\etal~\cite{mockus2000identifying}.  Lines modified or deleted on those big-fix commits are marked as buggy. Then we identified the original commits that introduce these bugs using SZZ algorithm~\cite{sliwerski2005changes}. Next, we used {\tt git blame} with {\tt --reverse}  option to locate those buggy lines in the buggy program version under investigation. Figure~\ref{fig:evolution} illustrates this process.  

Using this method, we found $1541$ additional buggy lines across all the versions of five \defj projects. Thus, in total, in this dataset, we studied $1761$ buggy lines, as shown in~\ref{tab:ssubj}.

\subsection{Evaluation Metric}
\label{sunsec:evaluation_metric}
\begin{figure}
\includegraphics[width=0.9\linewidth]{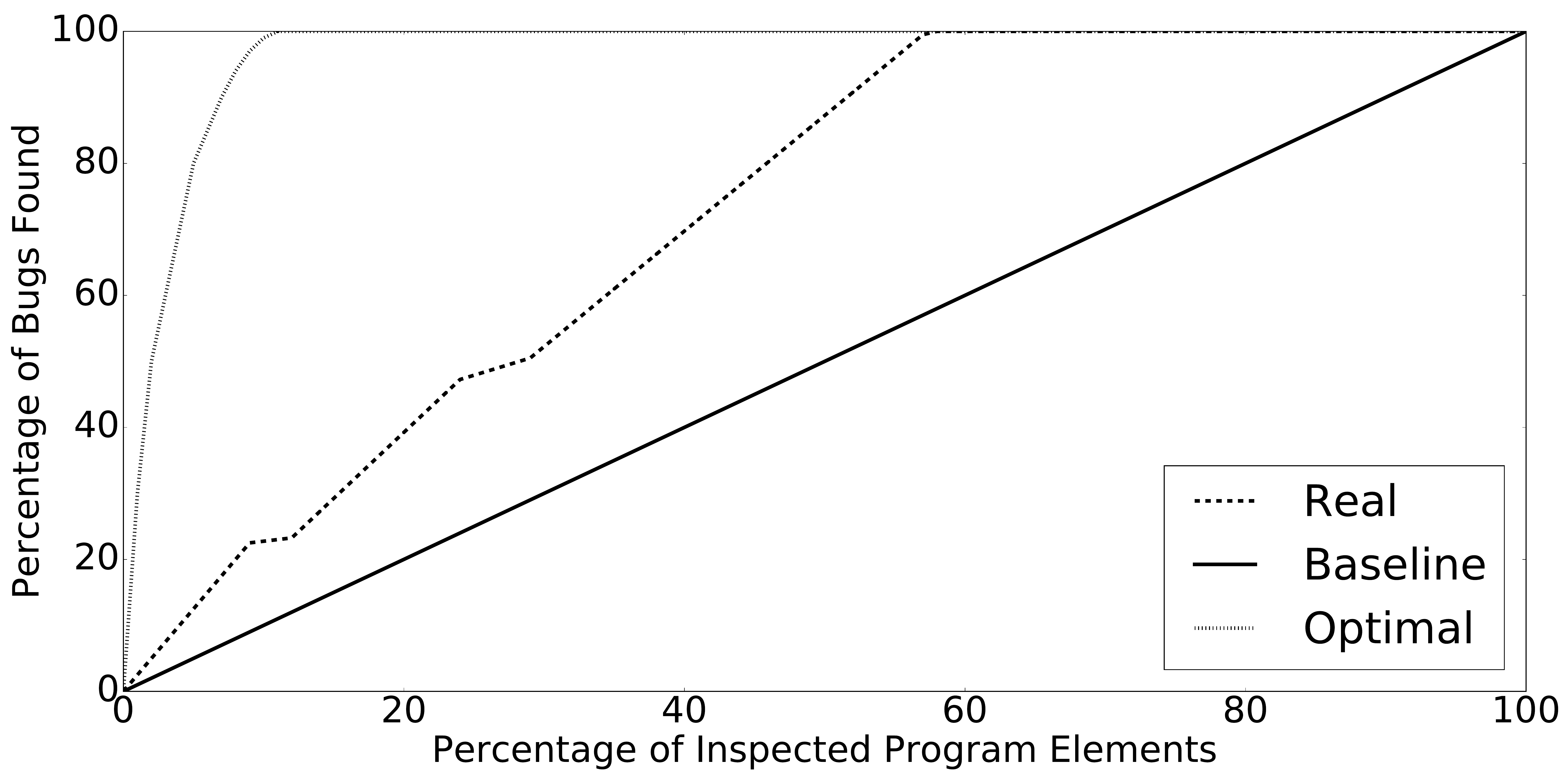}
\caption{{\small{\bf  Example Cost Effectiveness (CE) curve for bug localization. Baseline shows CE while inspecting random program elements. 
At optimal CE, 100\% of bugs are found when all the buggy program elemnets are  inspected first. A real CE falls somewhere in between baseline and optimal.}}}
\label{aucec_explain}
\end{figure}

To evaluate the bug localization capability of \tool, we adopted commonly used non-parametric measure from literature: Cost Effectiveness (CE) metric originally proposed by Arisholm~\etal~\cite{arisholm2010systematic} to investigate defects in telecom softwares. The main assumption behind this metric is, cost behind bug localization is the inspection effort\textemdash the number of Program Element (PE) needs to be inspected before locating the bug,  and the payoff is the percentage of bugs found. A Cost-Effectiveness(CE) curve shows percentage of inspected PE in {\it x-axis} and percentage of bugs found in {\it y-axis}. 

If bugs are uniformly distributed in the source code, 
by randomly inspecting $n\%$ of source PE, one might expect to find $n\%$ bugs.
The corresponding CE curve will be a straight line with $slope~\text{=}~1$ (see Figure~\ref{aucec_explain}). This is our baseline. Any ranking metric assigns suspiciousness score to each PE for bug localization. Then, PEs are inspected based on the decreasing order of suspiciousness score. An optimal ranking metric would assign scores in a way that all buggy PEs are ranked prior to the non-buggy PEs. So, inspecting top PEs would cover 100\% of the bugs.  For any real bug localization techniques, \eg Tarantula~\cite{jones2005tarantula}, Multric~\cite{xuan2014learning} etc., CE curve falls in between baseline and optimal.

AUCEC, the area of under the CE curve is a quantitative measurement describing how good a model is to find the bugs. Baseline AUCEC (random AUCEC) is 0.5. Optimal AUCEC would be very close to 1.00.  This AUCEC metric is a non-parametric metric similar to the ROC curve and does not depend on bug distribution~\cite{ray2016naturalness}, thus becomes standard in bug-localization literature~\cite{d2010extensive}. Higher AUCEC signifies higher prioritization of buggy lines over non-buggy lines and hence a better model.  For example, for the optimal CE, 100\% source program elements should not have to be inspected to find all the bugs; thus, optimal exhibits higher AUCEC than the baseline (see Figure~\ref{aucec_explain}). This intuition is the basis of our evaluation metric. 



\subsection{Implementation of \tool}
\label{subsec:impl}

We implemented \tool's learning to rank technique as described in Step 3 \& 4 of Section~\ref{sec:method} using two approaches. First, we used RankBoost~\cite{freund2003efficient} algorithm. RankBoost algorithm uses boosting ensemble technique to learn model parameter for ranking training instances. At each iteration, it learns one weak ranker and then re-weights the training data. At the final stage, it combines all the weak rankers to assign scores to the test data. This algorithm is used in the past by Xuan \etal~\cite{xuan2014learning} for implementing \sbfl at method level bug localization. Though there are many competitive approaches to implement \sbfl, Xuan \etal report the best results till date. Thus, we adapted their approach in \tool to locate bugs at line granularity. We used RankBoost implementation of standard {\bf {\it RankLib}}~\cite{ranklib} library for this purpose. There are two configurable parameters: $\beta$ (initial ranking metric) and $\gamma$ (number of neighbor). Following Xuan \etal, we set these two parameter values to Tarantula and 10 respectively. Table~\ref{tbl:compare_rankboost} shows the result. 

In the second approach, we used Random Forest Algorithm (RF) to implement the proposed learning to rank technique. 
Random Forest is an ensemble learning technique developed by Breiman~\cite{breiman2001random} based on a combination of a large set of decision trees. Each tree is trained by selecting a random set of features from a random data sample selected from the training corpus. In our case, the algorithm, therefore, chooses some \sbfl and/or entropy related features randomly in training phase (step 3).  RF then learns conditional probability distribution of the chosen features \wrt the bugginess of each line in the training dataset. In addition, RF learns the importances for different features for discrimination. 
During training, the model learns $M$ decision trees and corresponding probability distribution. In the testing phase, {\it suspiciousness} scores from each of the learned model, and calculate final {\it suspiciousness} score based on equation~\ref{eqn:ensemble}. For implementation, we used standard python {\tt scikit-learn} package~\cite{sklearn_api}.


\begin{center}
\begin{table}[!htpb]
\caption{\textbf{\small Performance (AUCEC$_{100}$) comparison of two learning to rank implementations that are used to realize \tool. Only \sbfl related features are used to establish a baseline.}}
\label{tbl:compare_rankboost}
\centering
\scriptsize
\begin{tabular}{l|r|r}
\toprule
\textbf{Project} & \textbf{RankBoost} & \textbf{Random Forest} \\
\midrule
\chart & 0.847 & 0.908\\
\closure & 0.797 & 0.894\\
\lang & 0.824 & 0.862\\
\maths & 0.864 & 0.876\\
\jtime & 0.846 & 0.918\\
Libtiff & 0.847 & 0.887\\
Lighttpd & 0.753 & 0.806\\
Php & 0.835 & 0.899\\
Python & 0.788 & 0.807\\
Wireshark  & 0.864 & 0.829\\
\bottomrule
\end{tabular}
\end{table}
\end{center}

We compare the performance of the above two approaches using AUCEC$_{100}$ score. For the comparison purpose, we only used \sbfl related features (\ie did not include entropy scores), since we first wanted to measure how the two approaches perform in traditional \sbfl setting. Table~\ref{tbl:compare_rankboost} reports the result. For all of the studied projects, except Wireshark, Random Forest is performing better. Thus, we carried out rest of our experiments using Random Forest based implementation, since this gives the best \sbfl performance at line level, even when we compare against state of the art Xuan \etal's technique.

\subsection{Research Questions}
\label{sec:rq}

To evaluate \tool, we investigate whether a good language model (\LM) that captures naturalness (hence unnaturalness) of a code element can improve spectrum based testing. Previously, Ray \etal~\cite{ray2016naturalness} and Wang \etal~\cite{wang2016bugram} demonstrated that unnaturalness of code elements (measured in terms of entropy) correlate with bugginess. Thus, \LM can help in bug localization in a static setting. In contrast, \sbfl is a dynamic approach that relies on the fact that code elements that are covered by more negative test cases are more bug-prone. Therefore, to understand the effectiveness of \tool, we will investigate whether the combination of the two can improve bug-localization as a whole. 

Since \LM based bug-localization approach says that more entropic code is more bug prone, and \sbfl says that code element covered by more negative test cases are more prone to bugs, to make the combined approach work, the difference between entropies of the buggy and non-buggy lines should be significant for the negative test spectra. Thus, to understand the potential of entropy,  we start our investigation with the following research question:

\RQ{rq1}{How is entropy associated with bugginess for different types of test spectra?}

If the answer to the above question is affirmative for failing test spectra,  entropy can be used along with \sbfl for bug-prediction.  For every code element, \sbfl provides suspiciousness scores, and  \LM predicts its uncertainty in terms of entropy. Thus, one may expect that among the lines with higher suspiciousness score, more entropic lines are even more likely to be buggy. We, therefore, investigate whether entropy can help improving the bug prediction capability of \tool over \sbfl.

\RQ{rq2}{Can entropy improve \sbfl's bug-localization capability?}


To build a good \LM based bug localization technique, we need a large code corpus with adequate bug history; this is often challenging for smaller projects. A similar problem arises for history based defect prediction models\textemdash for newer projects enough history is usually not available to build a good model. In such case, researchers, in general, rely on the evolutionary history of other projects~\cite{zimmermann2009cross}. To mitigate the threat of using our proposed approach for smaller code base, we leverage such cross-project defect prediction strategy. We investigate, whether a language model trained on different projects can still improve \sbfl's performance. 

\RQ{rq3}{What is the effect of entropy on \sbfl's bug localization capability in a cross-project setting?}

\section{Result}
\label{sec:result}
\setcounter{RQCounter}{0}

In this section, we answer the research questions introduced in~\ref{sec:rq}. Our investigation starts with whether the buggy lines in failing test spectrum are more entropic than non-buggy lines. Note that, all the buggy and non-buggy lines are annotated using the strategy described in~\ref{subsec:dc}.

\RQA{rq1a}{How is entropy associated with bugginess for different types of test spectra?}
\begin{figure}[!htpb]
  \centering
  \scriptsize
  \includegraphics[width=0.95\columnwidth]{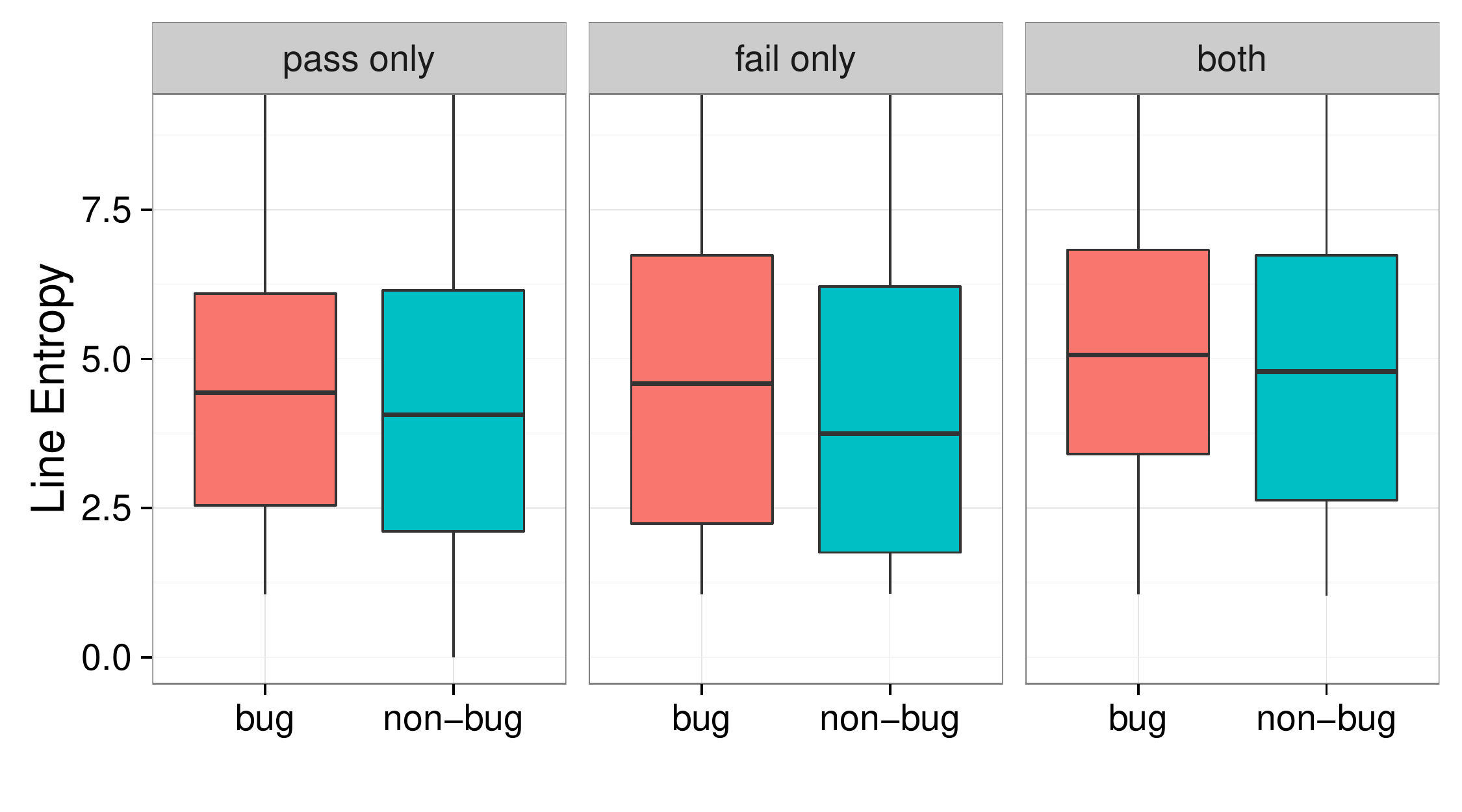}
   \begin{tabular}{l|c|lrl}
    \textbf{Type of }  &  \textbf{Percentage}  & \multicolumn{3}{c}{\textbf{Buggy Entropy > Non-buggy Entropy}} \\
    \textbf{test} & \textbf{buggy} & \textbf{Difference} &  &  \textbf{Effect Size}  \\
     \textbf{spectra} & \textbf{lines} & \textbf{(95\% conf interval)} & \textbf{p-value} &  \textbf{(Cohen's D)}  \\
    \toprule
    Fail only   & 4.22  &  0.06 to 0.99 & 0.027 &  0.20  ( small )  \\
    Pass only   & 0.85  &  0.00 to 0.28 & 0.051 &  0.06  ( negligible )  \\
    Both   & 1.60  &  0.09 to 0.38  & 0.002 &  0.09  ( negligible )  \\
    \bottomrule
    \end{tabular}%
  \caption{{\small{\bf For failing test spectra, buggy lines are more entropic than non-buggy lines; t-test confirms that the difference is statistically  significant (p-value $<$ 0.05) with small Cohen's D effect size. However, for other two spectra, the difference is negligible.}}}
  \label{fig:rq1}
\end{figure}


    
\begin{figure*}[!h]
\centering
\subfigure[\small{\bf Bug localization performance with an inspection budget of 20\% of total lines}]{
\centering
  \includegraphics[width=0.3\columnwidth]{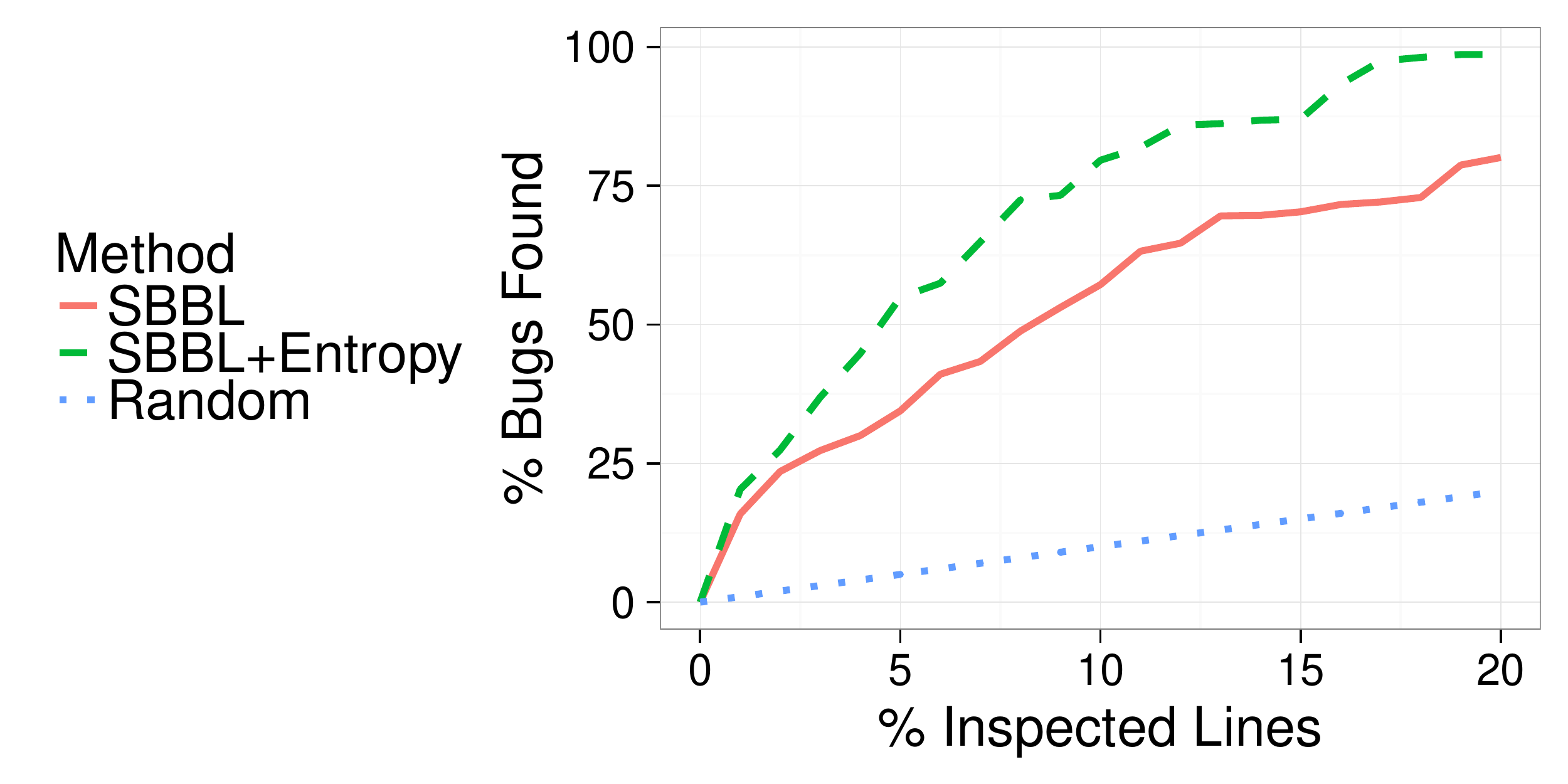}
\label{fig:rq2-summary}}
\quad
\subfigure[\small{\bf Bug localization performance with an inspection budget of 5\% of total lines for \defj dataset}]{
\centering
  \includegraphics[width=0.3\columnwidth]{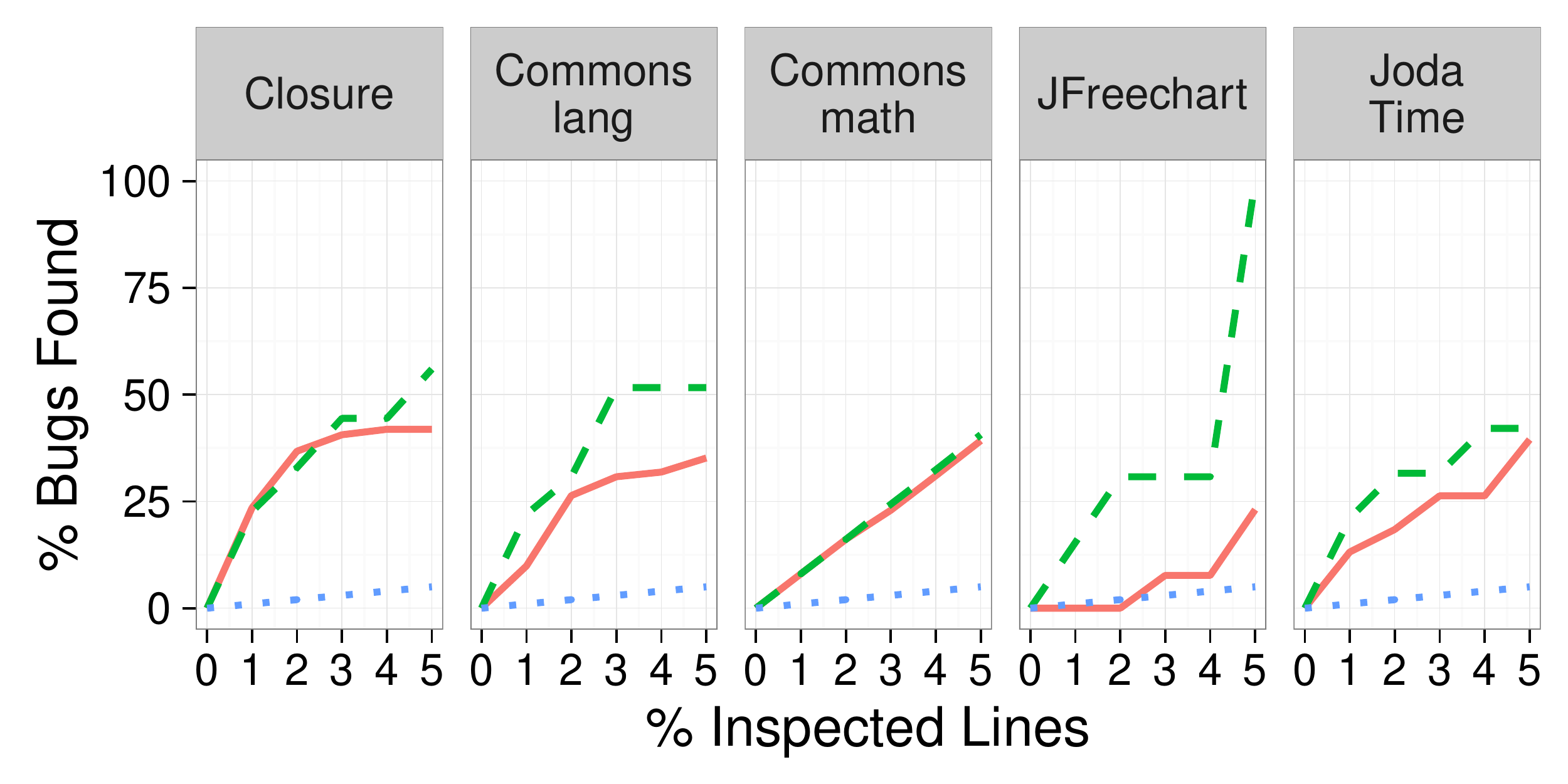}
\label{fig:rq2-defj}}
\quad
\subfigure[\small{\bf Bug localization performance with an inspection budget of 5\% of total lines for \mbug dataset}]
{
\centering
  \includegraphics[width=0.3\columnwidth]{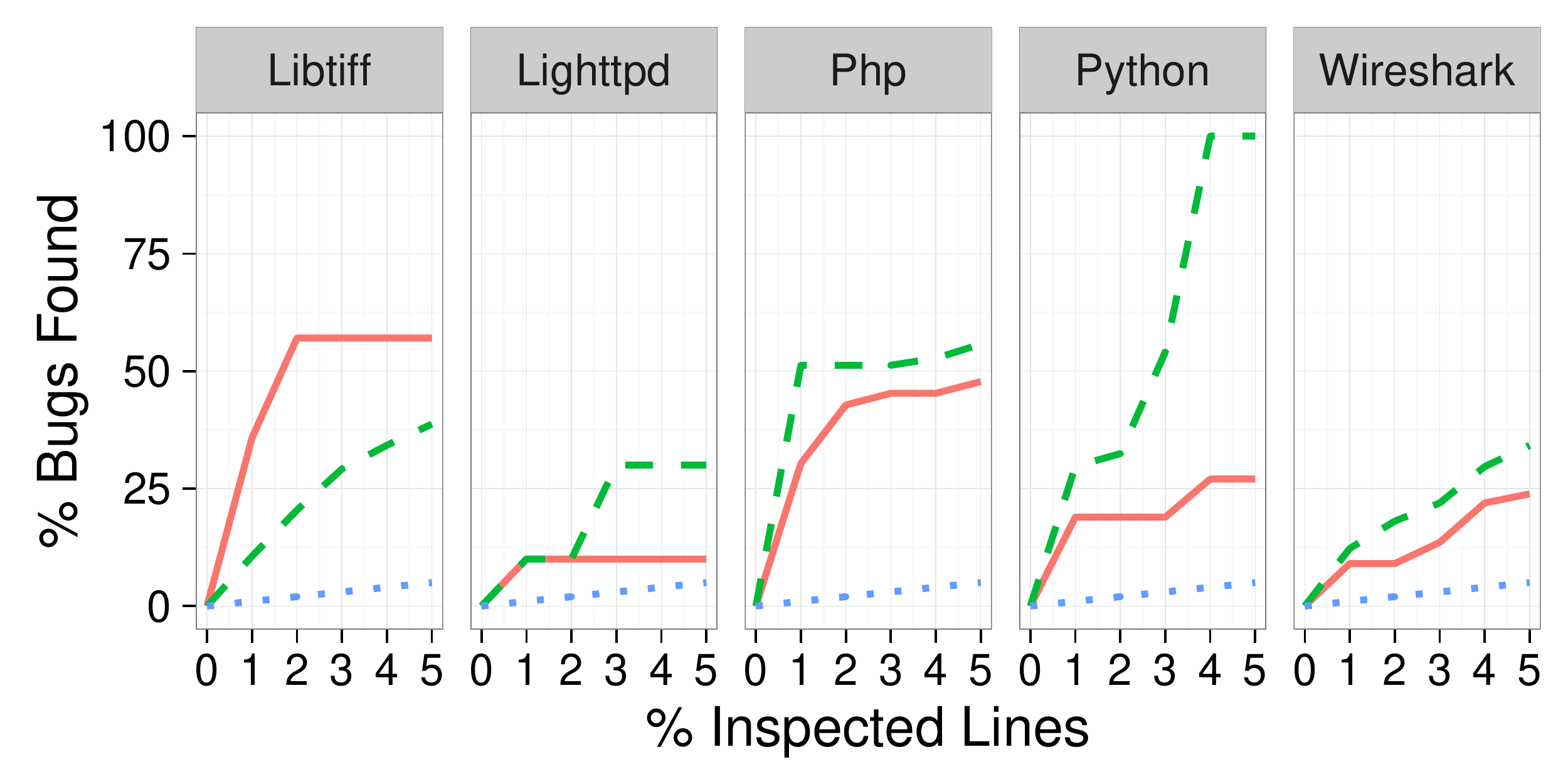}
\label{fig:rq2-mbug}}
\centering
\scriptsize
\begin{tabular}{l|c|c|c|l l l }
\toprule
{\bf Project} & {\bf AUCEC$_{100}$}  & \textbf{AUCEC$_{100}$}  & {\bf Gain} & \multicolumn{3}{c}{\bf Top Three Features}\\
& (\sbfl) & (\sbfl+Entropy)  &(\%)&Feature 1& Feature 2& Feature 3\\
\midrule
\chart	& 0.908 & 0.956 & 5.305 & Average Entropy  & Ochiai & RogersTanimoto\\
\closure	& 0.894 & 0.916 & 2.451 & Average Entropy  & Euclid & SimpleMatching\\
\lang	& 0.862	& 0.943	& 9.368 & Average Entropy  & SimpleMatching & RogersTanimoto\\
\maths	& 0.876	& 0.883 & 0.708 & Average Entropy  & Jaccard & RogersTanimoto\\
\jtime & 0.918 & 0.930	& 1.318 & Average Entropy  & Euclid & Hamming\\
Libtiff  & 0.887 & 0.904 & 2.004 & Average Entropy  & SimpleMatching & Hamming\\
Lighttpd & 0.806 & 0.943 & 17.016 & Average Entropy  & Jaccard & Ochiai \\
Php 	 & 0.899 & 0.941 & 4.675 & Average Entropy  & Hamming & Euclid\\
Python   & 0.807 & 0.961 & 19.133 & Average Entropy  & Sokal & SimpleMatching\\
Wireshark & 0.829 & 0.864 & 4.193 & Average Entropy  & Wong3 & Ample\\
\midrule
Average & 0.869 & 0.924 & 6.617\\  
\bottomrule
\end{tabular}
\caption{\textbf{\small Overall performance of \tool (\sbfl+ Entropy) vs.~\sbfl for localizing buggy lines.}}
\label{fig:rq2}
\end{figure*}

To answer this question, first, for each buggy version of \defj and \mbug dataset, we calculated the \ngm entropy value for every line in the project's source code. We also instrumented the source code to record the execution trace information at line level for passing and failing test cases. For \defj projects, we instrumented each file. Therefore, we recorded the execution traces for all files. However, \mbug projects were large in size (see~\ref{tab:ssubj}); hence due to technical limitations, we only instrumented the buggy files for this data set. Note that, this is not an unreasonable assumption; one can use other human-based or automated techniques, \eg  statistical or information retrieval based defect prediction, to locate potential buggy files~\cite{rahman2014comparing, walden2014predicting, saha2013improving}. Additionally, many of our buggy files have hundreds of lines of code, which is similar to the size of the code tested in previous \sbfl analyses~\cite{jones2002visualization}.

Next, we group the program lines based on the test execution traces:  
i)~\textit{Fail Only}: lines covered by only failing test cases (no passing test case exercise these lines), 
ii)~\textit{Pass Only}: lines covered by only passing test cases (no failing test cases exercised these lines), and 
iii)~\textit{Both}: lines covered by both failing and passing test cases. Note that, there is still a considerable number of program lines that are not covered by any test cases because the code coverage is not 100\% for any of the studied test suite. Such non-executed program lines are out of scope for this research question. For each group, we studied the \ngm entropy differences between buggy and non-buggy program lines. Figure~\ref{fig:rq1} shows the result.

Buggy lines, in general, have higher entropy than non-buggy lines in all three groups, as shown in the boxplot.  However, a t-test~\cite{zimmerman1987comparative} (see the Table below of~\ref{fig:rq1}) confirms that for {\em Fail Only} spectra,  the difference is statistically significant (p-value < 0.05), with a small Cohen's D effect size. Note that, a  similar effect size for overall buggy vs.~non-buggy entropy was also reported by previous studies~\cite{ray2016naturalness}. For {\em Pass Only} spectra, the difference is not statistically significant between buggy and non-buggy lines. Furthermore, although a small difference exists for {\em Both} spectra, the Cohen's D effect size is negligible. These results are also confirmed by Wilcoxon non-parametric test~\cite{mann1947test,zimmerman1987comparative}. 

In summary, we conclude that entropy is associated with bugginess for lines covered by failing test cases, while for the lines not covered by the failing test cases, entropy does not show significant association. This could be due to many factors such as the complexity or nuance of bugs that are not captured adequately by the test cases, or simply because of the quality of the test suite if it does not exercise many non-buggy lines which is low entropic. Certainly, more research is needed. Nevertheless, this result is significant for our study. Since, in failing spectra, entropy can further discriminate buggy and non-buggy lines, entropy score on top of the \sbfl based suspiciousness score would enhance the bug localization capability. Additionally, since entropy does not differentiate between buggy and non-buggy lines in the passing spectra, it will not boost \sbfl suspiciousness score for them.  Thus, we believe overall entropy would play an key role to improve \sbfl based bug localization  by increasing the suspiciousness scores of the buggy lines that cause tests to fail.  

\RS{rq1}{For failing test spectra, buggy lines, on average, are high entropic, \ie less probable, than non-buggy lines.}

So far, we have seen that entropy is associated with the bugginess of the code elements covered by failing test cases. Now, let us investigate whether \tool can leverage this property to localize bugs more effectively. 

\RQA{rq2a}{Can entropy improve \sbfl's bug-localization capability?}

To investigate this research question, we trained \tool's model in two settings:
i.~{\em \sbfl only}: 25 different \sbfl based suspiciousness scores, as shown in Table~\ref{tbl:rank_metrics} as features, and ii.~{\em \sbfl+~Entropy}: 25 different \sbfl based suspiciousness scores along with entropy related scores (see Section~\ref{sec:method}) as features, where entropies are generated by \ngmt language model (see equation~\ref{eqZscore}). We computed AUCEC at 100\% inspection budget for both the settings, \aucecsp and \aucecen respectively. The experiment is run 30 times per project and the final result is the average of those 30 runs. To understand relative improvements, we calculated relative percentage gain using equation~\ref{eqn:gain}.

\begin{equation}
\small
gain = \frac{AUCEC_{en} - AUCEC_{sp}}{AUCEC_{sp}} 
\label{eqn:gain}
\end{equation}
\begin{equation}
overall\_gain = \frac{\sum_{project}{AUCEC_{en} - AUCEC_{sp}}}{\sum_{project}{AUCEC_{sp}}} 
\label{eqn:overall_gain}
\end{equation}

\begin{figure*}[!h]
\centering
\subfigure[\small{\bf \defj dataset}]{
\centering
  \includegraphics[width=0.45\linewidth]{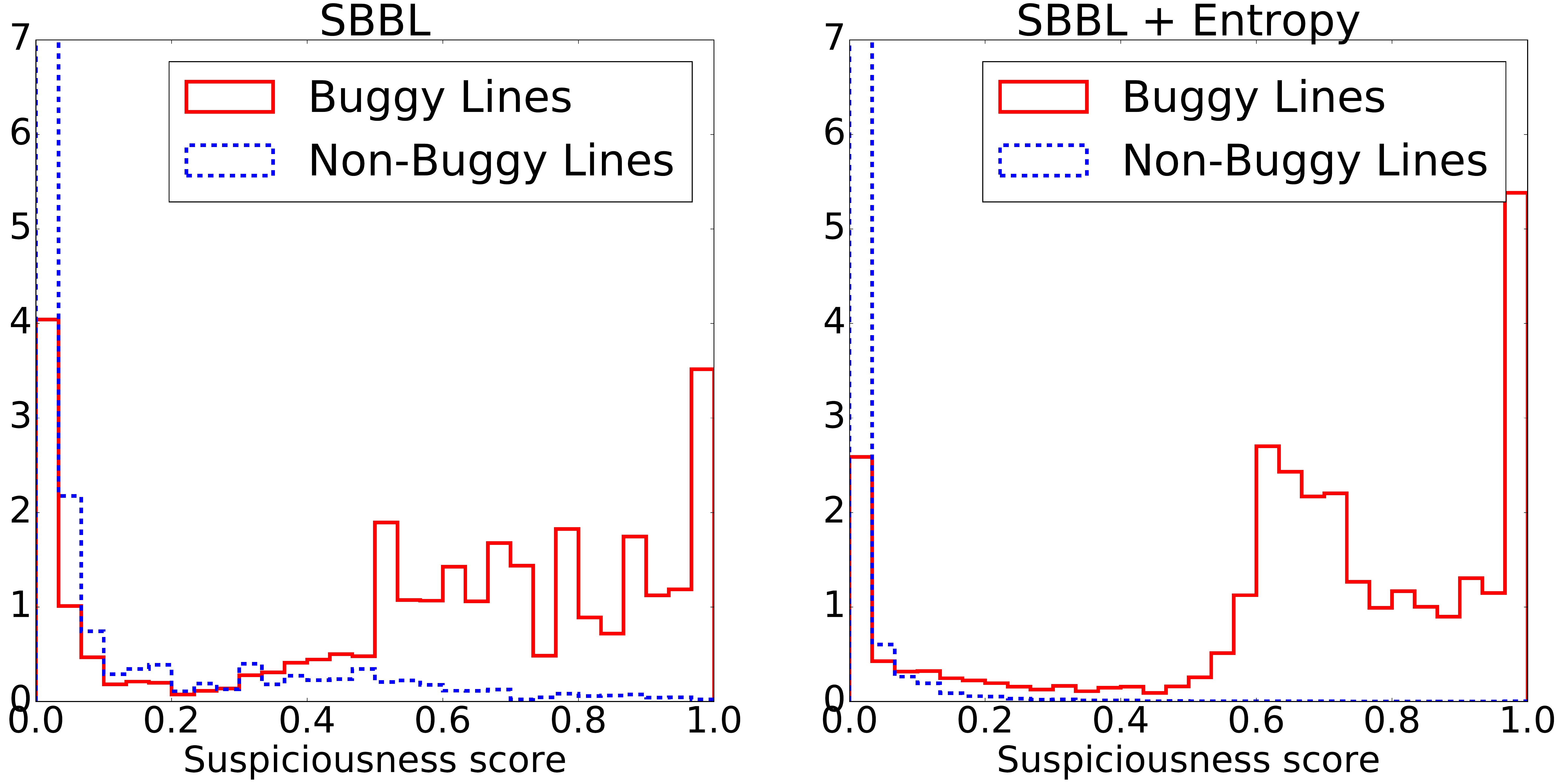}
\label{fig:defj_histogram}}
\quad
\subfigure[\small{\bf \mbug dataset}]{
\centering
  \includegraphics[width=0.45\linewidth]{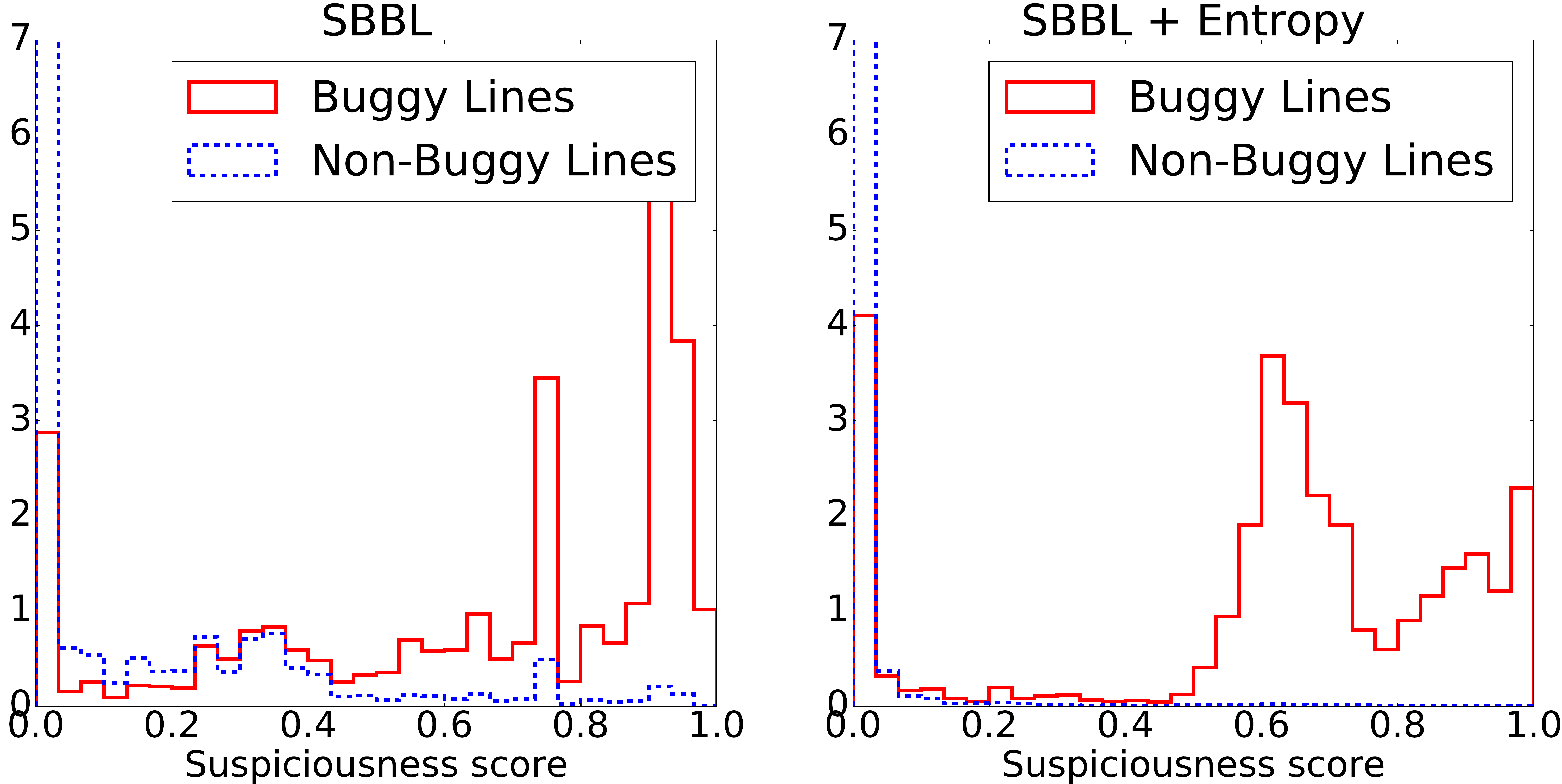}
\label{fig:mbug_histogram}}
\caption{\textbf{\small Histogram of suspiciousness score per source code lines. $x$-axis shows~\textit{suspiciousness} score given by equation~\ref{eqn:ensemble} and the $y$-axis is the relative frequency.}}
\label{fig:rq2-hist}
\end{figure*}

The results are shown in the bottom table of Figure~\ref{fig:rq2}. 
For all the projects we studied, AUCEC$_{100}$ increases when we incorporate entropy related features, with an overall gain of 6.617\% (3.81\% and 9.11\% for \defj and \mbug dataset respectively). The positive gain indicates that \LM can indeed improve the performance of \sbfl bug localization. Top three features (sorted based on their importance learned by Random Forest algorithm in the training phase) are also reported as Feature1, Feature2 and Feature3. For all the cases, Average Entropy is the most important feature. Thus, we conclude that, entropy plays the most important role in localizing bugs for \tool.

Next, we check how \tool performs with \sbfl+~entropy related features over \sbfl only features while inspecting small portion of source code lines (SLOC), as that is more realistic scenario. Figure~\ref{fig:rq2-summary} shows the overall effect of entropy across all the projects at an inspection budget of 20\%. With \sbfl only features, we can detect only 80.08\% of total buggy lines (AUCEC$_{20}$ is 0.10475). However, with \sbfl+~entropy features, we can detect 98.63\% buggy lines (AUCEC$_{20}$ is 0.13966). Thus we see an overall gain of 33.33\% in AUCEC$_{20}$ with entropy. 

Next, we check, at a even lower inception budget, how \tool performs. 
Figure~\ref{fig:rq2-defj} and Figure~\ref{fig:rq2-mbug} shows the cost-effectiveness curve for individual projects in \defj and \mbug dataset respectively. Here, \sbfl+~entropy yields better performance over \sbfl only setting for all the projects except two: for \maths both performs almost equal and for Libtiff, \tool performs worse. 

 We further investigate, why \LM is helping \sbfl to improve bug localization. Figure~\ref{fig:defj_histogram} and~\ref{fig:mbug_histogram} show the histogram of {\it suspiciousness} scores per line from \defj and \mbug dataset in both the settings. For \sbfl only setting, a large  proportion of actual non-buggy lines (marked in \blue{blue}) lie in the higher {\it suspiciousness} score range. So, those non-buggy lines are ranked in higher position than the buggy lines (marked in \Red{red}) having lower scores, lowering down the overall bug-localization performance. But, in \sbfl + Entropy setting,  the proportion of non-buggy lines having higher hybrid suspiciousness scores decreases, improving the overall bug localization performance. 

\RS{rq2}{Entropy, as derived by statistical language model, improved bug localization capabilities of \sbfl.}

For projects with smaller size and less number of test cases, \tool might not work well, since smaller projects do not have enough bug data to train \tool. To mitigate this threat, we now evaluate \tool's performance in cross-project settings. 

\begin{figure*}[!htpb]
\centering
\subfigure[\small{\bf Bug localization performance with an inspection budget of 20\% of total lines. 75.34\% and 54.5\% gain in the corresponding AUCEC values for \defj and \mbug projects respectively, when we use entropy.}]{
\centering
  \includegraphics[width=0.45\columnwidth]{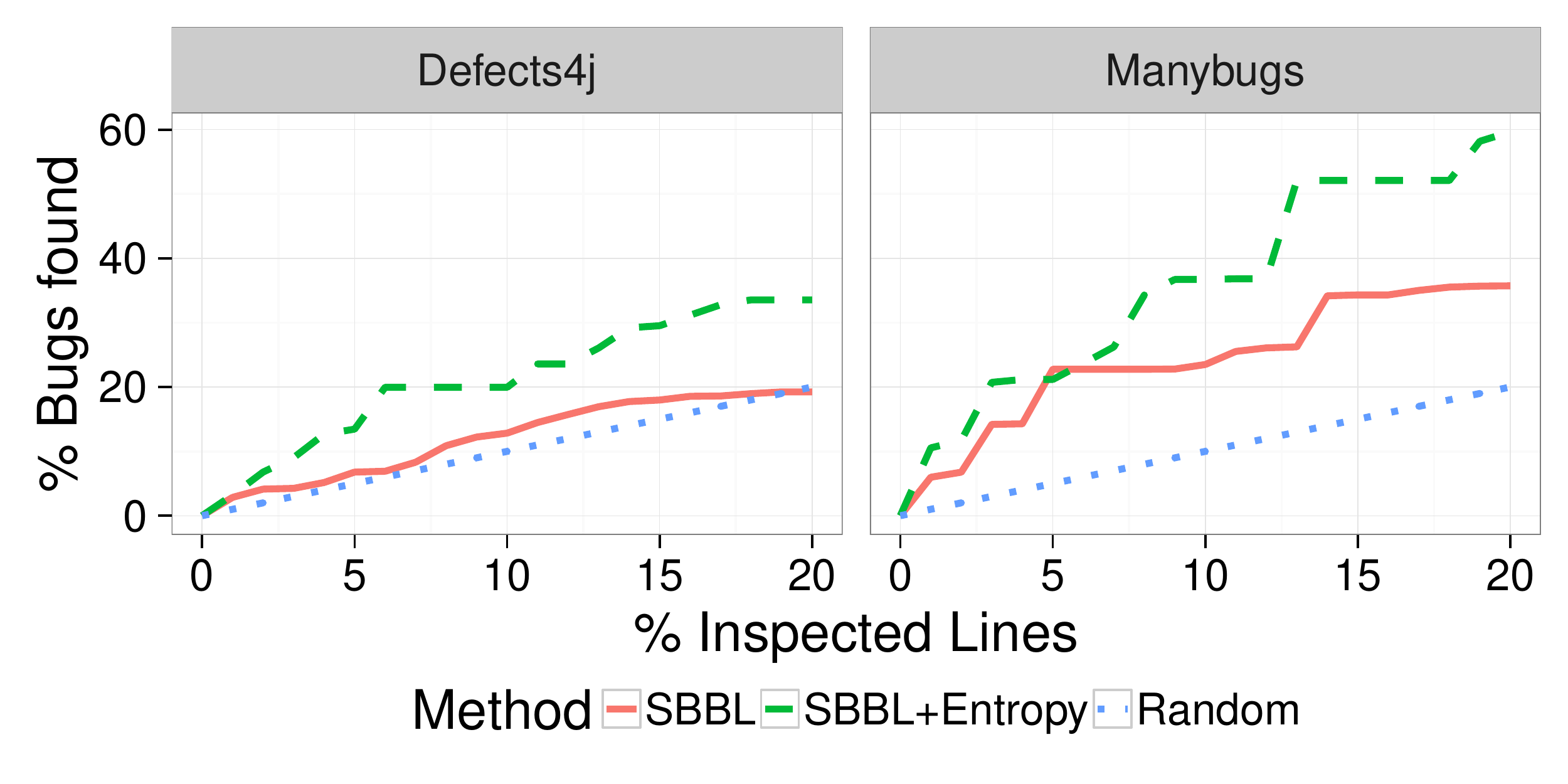}
\label{fig:rq3-summary_20}}
\quad
\subfigure[\small{\bf Bug localization performance with an inspection budget of 5\% of total lines. 94.16\% and 54.18\% gain in the corresponding AUCEC values for \defj and \mbug projects respectively, when we use entropy.}]{
\centering
  \includegraphics[width=0.45\columnwidth]{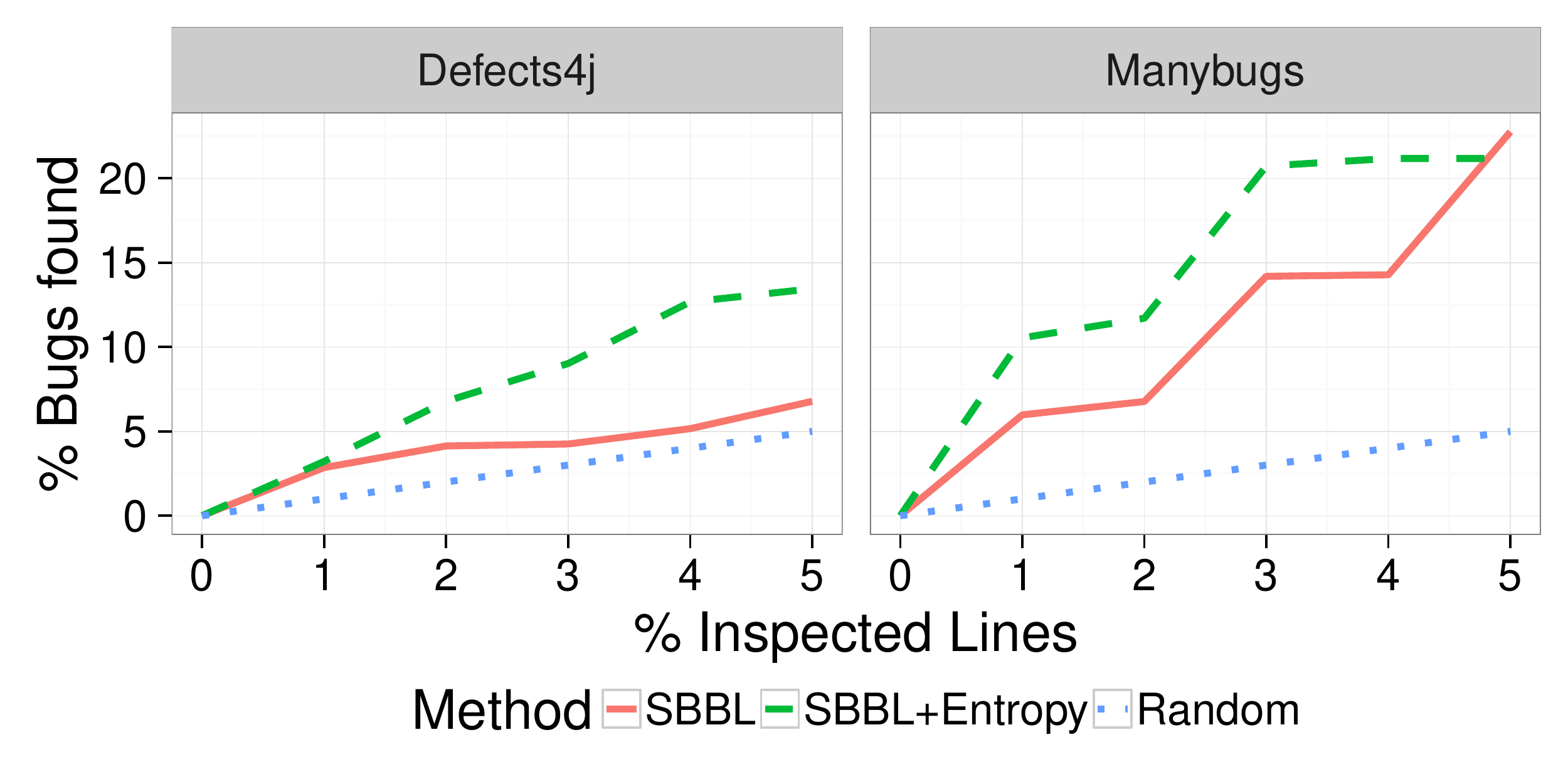}
\label{fig:rq3-summary_5}}
 \centering
 \scriptsize
\begin{tabular}{l|l|c|c|c|l l l}
\toprule
{\bf Dataset} & {\bf Project} & {\bf AUCEC$_{100}$}  & {\bf AUCEC$_{100}$}  & {\bf Overall Gain }  & \multicolumn{3}{c}{\bf Top Three Features}\\
&& (\sbfl only) & (\sbfl + entropy) & (\%) &Feature1&Feature2&Feature3\\
\midrule
&\chart & 0.878 & 0.955 & 8.882 & Average entropy  & Hamming & Euclid  \\
&\closure & 0.681 & 0.816 & 37.796 & Average entropy  & Ample & RusselRao \\
\defj &\lang & 0.619 &	0.840 & 36.655 & Average entropy  & Euclid & RogersTanimoto\\
&\maths & 0.743 & 0.774 &	4.172 & Mean entropy  & Euclid & Hamming\\
&\jtime & 0.662 & 0.794 & 19.939 & Average entropy  & Hamming & Sokal\\
\midrule
& Average & 0.717 & 0.836 & 16.59\\
\midrule
& Libtiff  & 0.565 & 0.729 & 45.026 & Average entropy  & Wong3 & RogersTanimoto\\
& Lighttpd & 0.616 & 0.848 & 37.623 & Average entropy  & RogersTanimoto & SimpleMatching\\
\mbug & Php 	 & 0.508 & 0.541 & 11.299 & Mean entropy  & Wong3 & Hamming\\
& Python   & 0.577 & 0.792 & 37.422 & Average entropy  & Ample & Euclid\\
& Wireshark & 0.757 & 0.898 & 20.489 & Average entropy  & RogersTanimoto & Sokal\\
\midrule
& Average & 0.605 & 0.762 & 25.95\\
\bottomrule
\end{tabular}
\caption{\textbf{\small Overall performance of \tool (\sbfl+ Entropy) vs.~\sbfl in cross project setting.}}
\label{fig:rq3}
\end{figure*}

\RQA{rq3a}{What is the effect of entropy on \sbfl's bug localization capability in a cross-project setting?}

To answer this research question, we first build a cross-project \LM.  To train the \LM for a given buggy project version, v$_{target}$, we select all the other buggy versions (v$_{train}$) from different projects with the following two constraints: (1) v$_{train}$ and  v$_{target}$ are written in the same language, and (2) v$_{train}$ are created before $v_{target}$.
The first constraint ensures that the \LM is trained on the same programming language specific features and usages.  Thus, for a given Java project, \LM only learns from other Java projects from \defj dataset. Similarly, for a C project, we choose other C projects from \mbug dataset for training.

 The second constraint emulates the real world scenario: to train a \LM for v$_{target}$, the training set (v$_{train}$) has to be available in the first place. Hence,  we organize different project versions of a dataset chronologically, based on their last commit date, as retrieved from version repository. Then, for v$_{target}$, we choose all the versions from different projects that are chronologically appeared before v$_{target}$. Thus, for both the dataset, we choose a buggy version as the target if there exists at least one previous buggy version from another project.

Next, for each v$_{target}$, we train \tool based on the features of v$_{train}$. Similar to our previous experiments, the features include entropy scores generated by \ngmt language model (see equation~\ref{eqZscore}), and 25 \sbfl specific suspiciousness scores,  for each program line. Based on the training, \tool then assigns a suspiciousness score per line in the test data, v$_{target}$, indicating the line's likelihood of being buggy.  Finally, we rank the lines with decreasing order of their suspiciousness score\textemdash line with the highest value tops the list.  To evaluate the performance of \tool at cross-project setting, we calculate the AUCEC$_n$ score, which basically tells us the rate and the percentages of bugs that can be found if a developer inspects n\% lines in the ranked-list returned by the tool.  We also repeat the experiment without entropy as a feature, \ie with only \sbfl suspiciousness score. As a baseline, we report the percentage of buggy lines found at an inspection budget of n, while we randomly choose n\% lines from source code. Figure~\ref{fig:rq3} shows the results.

Figure~\ref{fig:rq3-summary_20} shows that if developers inspect top 20\% of source code lines, following \sbfl+Entropy ranking scheme, 33.54\% and 59.65\% of the buggy lines are detected for \defj and \mbug respectively. In contrast, when we use only \sbfl based ranking scheme, 19.24\% and 35.72\% buggy lines can be detected for \defj and \mbug respectively. Thus, we see an overall gain of  75.34\% and 54.5\% of AUCEC$_{20}$, when we include entropy in the feature set. Note that, this gain even increases at a stricter inspection budget for \defj projects\textemdash Figure~\ref{fig:rq3-summary_5} reports an overall gain of 94.16\% in AUCEC$_{5}$. For \mbug, we see a gain of 54.18\%, which is similar to AUCEC$_{20}$ gain. Both ranking schemes perform better than random.   

The table at bottom in Figure~\ref{fig:rq3} shows AUCEC$_{100}$ values achieved by two strategies for each project at an inspection budget of 100. In both cases, developers would find 100\% of buggy lines since they inspect all the lines in the ranked list. However, the rate of identifying 100\% of buggy lines is higher using \tool than \sbfl only. Result shows that \tool has 16.59\% performance gain for \defj data set, and 25.95\% gain for \mbug, on average. As discussed earlier, including entropy as a feature benefits bug localization even more at lower inspection budget, which is a more realistic scenario.

We further check the relative importance of the features learned by \tool (last three columns of the Table in Figure~\ref{fig:rq3} lists the three most important features). Interestingly, in all the cases, entropy exhibits highest  importance than any \sbfl based features. The above results show that a good \LM can significantly improve \sbfl in a cross-project bug localization setting.

\RS{rq3}{Entropies derived from \LM significantly improves bug localization capabilities of \sbfl in a cross-project setting.}

\vspace{20pt}
\section{Related Work}
\label{sec:related}
Automatic \bug localization has been an active research area over two decades. Existing techniques can be broadly classified into two broad categories: i) static and ii) dynamic approaches. 

{\bf Static approaches} primarily rely on program source code. There are mainly two kinds of static approaches: a) program analysis based approaches, and  b) information retrieval (IR) based approaches. Program analysis based approaches  detect bugs by identifying well-known buggy patterns that frequently happened in practice. Therefore, although these approaches are effective in preventing bugs by enforcing good programming practices, they generally cannot detect functional bugs. {FindBugs}~\cite{ayewah2008using} is a popular example in this category. On the other hand, IR-based approaches, given a bug report, generally rank source code files based on the textual similarity between source code and the bug report so that potential buggy files ranked high in the ranked-list. These approaches are generally fast but identify bugs at coarse grained level.	 BugLocator~\cite{zhou2012should}, BLUiR~\cite{saha2013improving} are some of the examples in this category. 

There is a new line of work that recently started based on statistical modeling and machine learning. Wang et al.\cite{wang2016automatically} proposed a Deep Belief Network based approached to detect file level defects. Wang et al.\cite{wang2016bugram} used n-gram language model to generate a list of probable bugs.

{\bf Dynamic approaches} generally rely on the execution traces of test cases. \sbfl is a dynamic fault localization technique that leverages program spectra\textemdash program paths executed by passed and failed test cases~\cite{reps1997use} to compute a suspiciousness score of each program element. We have described \sbfl in detail in Section~\ref{subsec:spectrum}. 
Several metrics have been proposed in the literature to calculate the suspiciousness score. For example, Jones \etal presented Tarantula~\cite{jones2005tarantula} based on the fact that program elements executed by failed test cases are more likely to \bug than the elements not executed by them (see Table~\ref{tab:spectra}). Jaccard and Ochiai are some of the well known variants of this approach proposed by Abreu \etal~\cite{abreu2007ochiai_and_jaccard}. Xie \etal proposed five ranking metrics by theoretical analysis and four other metrics based on genetic algorithms~\cite{xie2013theoretical}. Later, Lucia \etal did a comprehensive study of the different ranking metrics and showed that no ranking metric is unanimously best~\cite{lucia2014extended}. \sbfl approaches can identify bugs at fine-grained level. 

Xuan \etal~\cite{xuan2014learning} proposed an approach to combine multiple ranking metrics. They adopted neighborhood based strategy to reduce the imbalance ratio of buggy and non-buggy program entities. For their algorithm, they need an initial metric to define the neighborhood. They sort the data based on an initial ranking metric. The filtered $\beta$ non-faulty entities before and after a faulty entity. After that, they applied state of the art "Learning to Rank" algorithms to combine all 25 suspiciousness scores. Their dependence on an initial ranking metric might cause a bias towards that metric. In contrast, to be unbiased to any of the metric, we considered all the data and applied state of the art random under-sampling\cite{batista2004study} technique. 

Gong \etal~\cite{gong2012interactive} proposed a feedback based fault localization system, which uses user feedback to improve performance. Pytlik \etal~\cite{pytlik2003automated} proposed the fault localization system using the likely invariant. Le et al.\cite{b2016learning} also proposed an approach similar to Pytlic et al. with a larger invariant set. Unlike this work, they experimented on method level fault localization system. Sahoo et al. extended Pytlik et al's work. Their work is on test case generation and also they adopted backward slicing to reduce the number of program element to be considered. 

{\bf Multi-modal techniques} generally combine two or more model of \bug localization to improve the accuracy further. Le et al.~\cite{le2015information} proposed a multi-modal technique for bug localization that basically combines the IR and spectrum based bug localization together. Their technique needs three artifacts: i) a bug report, ii) program source code, and iii) a set of test cases having at least a fault reproducing test case. Their technique first rank the source code methods based on the textual similarity between bug report and source code methods. Then using program spectra, they rank the source code lines and also identifies a list of suspicious words associated with the bug. Finally they combine these scores using a probabilistic model which is trained on a set of previously fixed bugs. Based on an empirical evaluation on 157 real bugs from four software systems, their model outperforms a state-of-the-art IR-based bug localization technique, a state-of-the-art spectrum-based bug localization technique, and three state-of-the-art multi-modal feature location methods that are adapted for bug localization. 

The proposed approach in this paper is also multi-modal in nature. However, instead of combining IR-based textual similarity score,inspired by Ray et al.'s\cite{ray2016naturalness} finding that the buggy codes are unnatural, and thus entropy of a buggy source code is naturally high, we combine source code entropy with program spectra to improve bug localization. To our knowledge, no one leveraged the localness of code and test spectrum together to locate faults. The advantage of our approach is that we do not need any bug report which may not be available for development bugs. Therefore, our approach is complementary to Le et al's approach.

\section{Threats to Validity}
\label{sec:threats}

Efficiency of \tool depends on the availabity of previous bugs on which the model will be trained. To minimize this threat, we demonstrated that \tool works well in cross-project setting.

\tool is also dependent on the adequecy of the test suite. If there are not enough failing test cases, performance of \sbfl may get hurt, and hence \tool's performance will also be worse. However, since \tool is a hybrid approach, it does not solely depend on test suites. The \LM based part will still be able to locate bugs since the latter does not require anything but source code.

 Further, to annotate buggy lines, we rely on the publicly available bug-dataset and some evolutionary bugs. It can be possible there are other bugs lying in the code corpus that are polluting our results. However, at any given point of time it is impossible to know the presence of all the bugs in a software.

 Finally, to minimize threats due to external validity, we evaluated \tool on 10 projects for 2 languages: C and Java. This proves, \tool is not restricted to any particular programming language. 

\section{Conclusion}
\label{sec:conc}
While spectrum based bug localization is an extensively studied research area, studying buggy code in association with code naturalness (thus unnaturalness) is relatively new. In this work, we introduced the notion of code entropy as captured by statistical language model in \sbfl to make the overall bug localization more robust, and proposed an effective way of integrating entropy with \sbfl suspicious scores. We implemented our concept in a prototype called \tool. Our experimental results with \tool show that code entropy is positively correlated with the buggy lines executed by the failing test cases. Our results also demonstrate that \tool, when configured to use both  entropy and \sbfl, outperforms the configuration that uses only various \sbfl as features. \tool can also be leveraged for detecting bugs in cross-project setting for relatively new projects, where project bug database and evolutionary history is not strong enough. 

Our future direction includes  leveraging \tool to repair buggy program line more effectively, and improving \tool further by incorporating language model which captures not only the syntactic structure but also code semantic structure. 




\balance
\bibliographystyle{abbrv}
\bibliography{main} 

\begin{thebibliography}{10}

\bibitem{ranklib}
Ranklib (\url{https://sourceforge.net/p/lemur/wiki/RankLib/}).
\newblock \url{https://sourceforge.net/p/lemur/wiki/RankLib/}.

\bibitem{abreu2009practical}
R.~Abreu, P.~Zoeteweij, R.~Golsteijn, and A.~J. Van~Gemund.
\newblock A practical evaluation of spectrum-based fault localization.
\newblock {\em Journal of Systems and Software}, 82(11):1780--1792, 2009.

\bibitem{abreu2007ochiai_and_jaccard}
R.~Abreu, P.~Zoeteweij, and A.~J. Van~Gemund.
\newblock On the accuracy of spectrum-based fault localization.
\newblock In {\em Testing: Academic and Industrial Conference Practice and
  Research Techniques-MUTATION, 2007. TAICPART-MUTATION 2007}, pages 89--98.
  IEEE, 2007.

\bibitem{arisholm2010systematic}
E.~Arisholm, L.~C. Briand, and E.~B. Johannessen.
\newblock A systematic and comprehensive investigation of methods to build and
  evaluate fault prediction models.
\newblock {\em Journal of Systems and Software}, 83(1):2--17, 2010.

\bibitem{ayewah2008using}
N.~Ayewah, D.~Hovemeyer, J.~D. Morgenthaler, J.~Penix, and W.~Pugh.
\newblock Using static analysis to find bugs.
\newblock {\em IEEE software}, 25(5), 2008.

\bibitem{b2016learning}
T.-D. B~Le, D.~Lo, C.~Le~Goues, and L.~Grunske.
\newblock A learning-to-rank based fault localization approach using likely
  invariants.
\newblock In {\em Proceedings of the 25th International Symposium on Software
  Testing and Analysis}, pages 177--188. ACM, 2016.

\bibitem{batista2004study}
G.~E. Batista, R.~C. Prati, and M.~C. Monard.
\newblock A study of the behavior of several methods for balancing machine
  learning training data.
\newblock {\em ACM Sigkdd Explorations Newsletter}, 6(1):20--29, 2004.

\bibitem{breiman2001random}
L.~Breiman.
\newblock Random forests.
\newblock {\em Machine learning}, 45(1):5--32, 2001.

\bibitem{brown1992class_ngram}
P.~F. Brown, P.~V. Desouza, R.~L. Mercer, V.~J.~D. Pietra, and J.~C. Lai.
\newblock Class-based n-gram models of natural language.
\newblock {\em Computational linguistics}, 18(4):467--479, 1992.

\bibitem{sklearn_api}
L.~Buitinck, G.~Louppe, M.~Blondel, F.~Pedregosa, A.~Mueller, O.~Grisel,
  V.~Niculae, P.~Prettenhofer, A.~Gramfort, J.~Grobler, R.~Layton,
  J.~VanderPlas, A.~Joly, B.~Holt, and G.~Varoquaux.
\newblock {API} design for machine learning software: experiences from the
  scikit-learn project.
\newblock In {\em ECML PKDD Workshop: Languages for Data Mining and Machine
  Learning}, pages 108--122, 2013.

\bibitem{campbell2014syntax}
J.~C. Campbell, A.~Hindle, and J.~N. Amaral.
\newblock Syntax errors just aren't natural: improving error reporting with
  language models.
\newblock In {\em Proceedings of the 11th Working Conference on Mining Software
  Repositories}, pages 252--261. ACM, 2014.

\bibitem{chawla2004editorial}
N.~V. Chawla, N.~Japkowicz, and A.~Kotcz.
\newblock Editorial: special issue on learning from imbalanced data sets.
\newblock {\em ACM Sigkdd Explorations Newsletter}, 6(1):1--6, 2004.

\bibitem{Chelf:2002:Paste}
B.~Chelf, D.~Engler, and S.~Hallem.
\newblock How to write system-specific, static checkers in metal.
\newblock In {\em Proceedings of the 2002 ACM SIGPLAN-SIGSOFT Workshop on
  Program Analysis for Software Tools and Engineering}, PASTE '02, pages
  51--60, New York, NY, USA, 2002. ACM.

\bibitem{cleve2005locating}
H.~Cleve and A.~Zeller.
\newblock Locating causes of program failures.
\newblock In {\em Software Engineering, 2005. ICSE 2005. Proceedings. 27th
  International Conference on}, pages 342--351. IEEE, 2005.

\bibitem{copeland2005pmd}
T.~Copeland.
\newblock {\em PMD applied}.
\newblock Centennial Books San Francisco, 2005.

\bibitem{d2010extensive}
M.~D'Ambros, M.~Lanza, and R.~Robbes.
\newblock An extensive comparison of bug prediction approaches.
\newblock In {\em Mining Software Repositories (MSR), 2010 7th IEEE Working
  Conference on}, pages 31--41. IEEE, 2010.

\bibitem{dietterich2002ensemble}
T.~G. Dietterich.
\newblock Ensemble learning.
\newblock {\em The handbook of brain theory and neural networks}, 2:110--125,
  2002.

\bibitem{Engler:2001:SOSP}
D.~Engler, D.~Y. Chen, S.~Hallem, A.~Chou, and B.~Chelf.
\newblock Bugs as deviant behavior: A general approach to inferring errors in
  systems code.
\newblock In {\em Proceedings of the Eighteenth ACM Symposium on Operating
  Systems Principles}, SOSP '01, pages 57--72, New York, NY, USA, 2001. ACM.

\bibitem{findbugs}
FindBugs.
\newblock \url{http://findbugs.sourceforge.net/}.
\newblock Accessed 2015/03/10.

\bibitem{Franks:2015:ICSE}
C.~Franks, Z.~Tu, P.~Devanbu, and V.~Hellendoorn.
\newblock Cacheca: A cache language model based code suggestion tool.
\newblock In {\em ICSE Demonstration Track}, 2015.

\bibitem{freund2003efficient}
Y.~Freund, R.~Iyer, R.~E. Schapire, and Y.~Singer.
\newblock An efficient boosting algorithm for combining preferences.
\newblock {\em Journal of machine learning research}, 4(Nov):933--969, 2003.

\bibitem{gabel2010study}
M.~Gabel and Z.~Su.
\newblock A study of the uniqueness of source code.
\newblock In {\em Proceedings of the eighteenth ACM SIGSOFT international
  symposium on Foundations of software engineering}, pages 147--156. ACM, 2010.

\bibitem{gong2012interactive}
L.~Gong, D.~Lo, L.~Jiang, and H.~Zhang.
\newblock Interactive fault localization leveraging simple user feedback.
\newblock In {\em Software Maintenance (ICSM), 2012 28th IEEE International
  Conference on}, pages 67--76. IEEE, 2012.

\bibitem{gunneflo1989evaluation}
U.~Gunneflo, J.~Karlsson, and J.~Torin.
\newblock Evaluation of error detection schemes using fault injection by
  heavy-ion radiation.
\newblock In {\em Fault-Tolerant Computing, 1989. FTCS-19. Digest of Papers.,
  Nineteenth International Symposium on}, pages 340--347. IEEE, 1989.

\bibitem{hellendoorn2015will}
V.~J. Hellendoorn, P.~T. Devanbu, and A.~Bacchelli.
\newblock Will they like this?: evaluating code contributions with language
  models.
\newblock In {\em Proceedings of the 12th Working Conference on Mining Software
  Repositories}, pages 157--167. IEEE Press, 2015.

\bibitem{Hindle:2012:ICSE}
A.~Hindle, E.~Barr, M.~Gabel, Z.~Su, and P.~Devanbu.
\newblock {On the naturalness of software}.
\newblock In {\em ICSE}, pages 837--847, 2012.

\bibitem{hindle2012naturalness}
A.~Hindle, E.~T. Barr, Z.~Su, M.~Gabel, and P.~Devanbu.
\newblock On the naturalness of software.
\newblock In {\em 2012 34th International Conference on Software Engineering
  (ICSE)}, pages 837--847. IEEE, 2012.

\bibitem{japkowicz2002class}
N.~Japkowicz and S.~Stephen.
\newblock The class imbalance problem: A systematic study.
\newblock {\em Intelligent data analysis}, 6(5):429--449, 2002.

\bibitem{johnson2013don}
B.~Johnson, Y.~Song, E.~Murphy-Hill, and R.~Bowdidge.
\newblock Why don't software developers use static analysis tools to find bugs?
\newblock In {\em Software Engineering (ICSE), 2013 35th International
  Conference on}, pages 672--681. IEEE, 2013.

\bibitem{jones2005tarantula}
J.~A. Jones and M.~J. Harrold.
\newblock Empirical evaluation of the tarantula automatic fault-localization
  technique.
\newblock In {\em Proceedings of the 20th IEEE/ACM international Conference on
  Automated software engineering}, pages 273--282. ACM, 2005.

\bibitem{jones2002visualization}
J.~A. Jones, M.~J. Harrold, and J.~Stasko.
\newblock Visualization of test information to assist fault localization.
\newblock In {\em Proceedings of the 24th international conference on Software
  engineering}, pages 467--477. ACM, 2002.

\bibitem{just2014defects4j}
R.~Just, D.~Jalali, and M.~D. Ernst.
\newblock Defects4j: A database of existing faults to enable controlled testing
  studies for java programs.
\newblock In {\em Proceedings of the 2014 International Symposium on Software
  Testing and Analysis}, pages 437--440. ACM, 2014.

\bibitem{le2015information}
T.-D.~B. Le, R.~J. Oentaryo, and D.~Lo.
\newblock Information retrieval and spectrum based bug localization: better
  together.
\newblock In {\em Proceedings of the 2015 10th Joint Meeting on Foundations of
  Software Engineering}, pages 579--590. ACM, 2015.

\bibitem{le2015manybugs}
C.~Le~Goues, N.~Holtschulte, E.~K. Smith, Y.~Brun, P.~Devanbu, S.~Forrest, and
  W.~Weimer.
\newblock The manybugs and introclass benchmarks for automated repair of c
  programs.
\newblock {\em IEEE Transactions on Software Engineering}, 41(12):1236--1256,
  2015.

\bibitem{le2012representations}
C.~Le~Goues, W.~Weimer, and S.~Forrest.
\newblock Representations and operators for improving evolutionary software
  repair.
\newblock In {\em Proceedings of the 14th annual conference on Genetic and
  evolutionary computation}, pages 959--966. ACM, 2012.

\bibitem{li2007mcrank}
P.~Li, C.~J. Burges, Q.~Wu, J.~Platt, D.~Koller, Y.~Singer, and S.~Roweis.
\newblock Mcrank: Learning to rank using multiple classification and gradient
  boosting.
\newblock In {\em NIPS}, volume~7, pages 845--852, 2007.

\bibitem{liblit2005scalable}
B.~Liblit, M.~Naik, A.~X. Zheng, A.~Aiken, and M.~I. Jordan.
\newblock Scalable statistical bug isolation.
\newblock {\em ACM SIGPLAN Notices}, 40(6):15--26, 2005.

\bibitem{liu2005sober}
C.~Liu, X.~Yan, L.~Fei, J.~Han, and S.~P. Midkiff.
\newblock Sober: statistical model-based bug localization.
\newblock In {\em ACM SIGSOFT Software Engineering Notes}, volume~30, pages
  286--295. ACM, 2005.

\bibitem{lucia2014extended}
L.~Lucia, D.~Lo, L.~Jiang, F.~Thung, and A.~Budi.
\newblock Extended comprehensive study of association measures for fault
  localization.
\newblock {\em Journal of Software: Evolution and Process}, 26(2):172--219,
  2014.

\bibitem{mann1947test}
H.~B. Mann and D.~R. Whitney.
\newblock On a test of whether one of two random variables is stochastically
  larger than the other.
\newblock {\em The annals of mathematical statistics}, pages 50--60, 1947.

\bibitem{mockus2000identifying}
A.~Mockus and L.~G. Votta.
\newblock Identifying reasons for software changes using historic databases.
\newblock In {\em icsm}, pages 120--130, 2000.

\bibitem{pytlik2003automated}
B.~Pytlik, M.~Renieris, S.~Krishnamurthi, and S.~P. Reiss.
\newblock Automated fault localization using potential invariants.
\newblock {\em arXiv preprint cs/0310040}, 2003.

\bibitem{rahman2014comparing}
F.~Rahman, S.~Khatri, E.~T. Barr, and P.~Devanbu.
\newblock Comparing static bug finders and statistical prediction.
\newblock In {\em Proceedings of the 36th International Conference on Software
  Engineering}, pages 424--434. ACM, 2014.

\bibitem{rao2011retrieval}
S.~Rao and A.~Kak.
\newblock Retrieval from software libraries for bug localization: a comparative
  study of generic and composite text models.
\newblock In {\em Proceedings of the 8th Working Conference on Mining Software
  Repositories}, pages 43--52. ACM, 2011.

\bibitem{ray2016naturalness}
B.~Ray, V.~Hellendoorn, S.~Godhane, Z.~Tu, A.~Bacchelli, and P.~Devanbu.
\newblock On the naturalness of buggy code.
\newblock In {\em Proceedings of the 38th International Conference on Software
  Engineering}, pages 428--439. ACM, 2016.

\bibitem{Raychev:2014:PLDI}
V.~Raychev, M.~Vechev, and E.~Yahav.
\newblock Code completion with statistical language models.
\newblock In {\em PLDI}, pages 419--428, 2014.

\bibitem{reps1997use}
T.~Reps, T.~Ball, M.~Das, and J.~Larus.
\newblock The use of program profiling for software maintenance with
  applications to the year 2000 problem.
\newblock In {\em Software Engineering—ESEC/FSE'97}, pages 432--449.
  Springer, 1997.

\bibitem{saha2013improving}
R.~K. Saha, M.~Lease, S.~Khurshid, and D.~E. Perry.
\newblock Improving bug localization using structured information retrieval.
\newblock In {\em Automated Software Engineering (ASE), 2013 IEEE/ACM 28th
  International Conference on}, pages 345--355. IEEE, 2013.

\bibitem{segall1995fiat}
Z.~Segall, D.~Vrsalovic, D.~Siewiorek, D.~Ysskin, J.~Kownacki, J.~Barton,
  R.~Dancey, A.~Robinson, and T.~Lin.
\newblock Fiat-fault injection based automated testing environment.
\newblock In {\em Fault-Tolerant Computing, 1995, Highlights from Twenty-Five
  Years., Twenty-Fifth International Symposium on}, page 394. IEEE, 1995.

\bibitem{sliwerski2005changes}
J.~{\'S}liwerski, T.~Zimmermann, and A.~Zeller.
\newblock When do changes induce fixes?
\newblock In {\em ACM sigsoft software engineering notes}, volume~30, pages
  1--5. ACM, 2005.

\bibitem{Tu:2014:FSE}
Z.~Tu, Z.~Su, and P.~Devanbu.
\newblock On the localness of software.
\newblock In {\em SIGSOFT FSE}, pages 269--280, 2014.

\bibitem{tu2014localness}
Z.~Tu, Z.~Su, and P.~Devanbu.
\newblock On the localness of software.
\newblock In {\em Proceedings of the 22nd ACM SIGSOFT International Symposium
  on Foundations of Software Engineering}, pages 269--280. ACM, 2014.

\bibitem{walden2014predicting}
J.~Walden, J.~Stuckman, and R.~Scandariato.
\newblock Predicting vulnerable components: Software metrics vs text mining.
\newblock In {\em Software Reliability Engineering (ISSRE), 2014 IEEE 25th
  International Symposium on}, pages 23--33. IEEE, 2014.

\bibitem{wang2016bugram}
S.~Wang, D.~Chollak, D.~Movshovitz-Attias, and L.~Tan.
\newblock Bugram: bug detection with n-gram language models.
\newblock In {\em Proceedings of the 31st IEEE/ACM International Conference on
  Automated Software Engineering}, pages 708--719. ACM, 2016.

\bibitem{wang2016automatically}
S.~Wang, T.~Liu, and L.~Tan.
\newblock Automatically learning semantic features for defect prediction.
\newblock In {\em Proceedings of the 38th International Conference on Software
  Engineering}, pages 297--308. ACM, 2016.

\bibitem{xie2013theoretical}
X.~Xie, T.~Y. Chen, F.-C. Kuo, and B.~Xu.
\newblock A theoretical analysis of the risk evaluation formulas for
  spectrum-based fault localization.
\newblock {\em ACM Transactions on Software Engineering and Methodology
  (TOSEM)}, 22(4):31, 2013.

\bibitem{xuan2014learning}
J.~Xuan and M.~Monperrus.
\newblock Learning to combine multiple ranking metrics for fault localization.
\newblock In {\em ICSME-30th International Conference on Software Maintenance
  and Evolution}, 2014.

\bibitem{ye2014learning}
X.~Ye, R.~Bunescu, and C.~Liu.
\newblock Learning to rank relevant files for bug reports using domain
  knowledge.
\newblock In {\em Proceedings of the 22nd ACM SIGSOFT International Symposium
  on Foundations of Software Engineering}, pages 689--699. ACM, 2014.

\bibitem{zeller2002simplifying}
A.~Zeller and R.~Hildebrandt.
\newblock Simplifying and isolating failure-inducing input.
\newblock {\em IEEE Transactions on Software Engineering}, 28(2):183--200,
  2002.

\bibitem{zhou2012should}
J.~Zhou, H.~Zhang, and D.~Lo.
\newblock Where should the bugs be fixed?-more accurate information
  retrieval-based bug localization based on bug reports.
\newblock In {\em Proceedings of the 34th International Conference on Software
  Engineering}, pages 14--24. IEEE Press, 2012.

\bibitem{zimmerman1987comparative}
D.~W. Zimmerman.
\newblock Comparative power of student t test and mann-whitney u test for
  unequal sample sizes and variances.
\newblock {\em The Journal of Experimental Education}, 55(3):171--174, 1987.

\bibitem{zimmermann2009cross}
T.~Zimmermann, N.~Nagappan, H.~Gall, E.~Giger, and B.~Murphy.
\newblock Cross-project defect prediction: a large scale experiment on data vs.
  domain vs. process.
\newblock In {\em Proceedings of the the 7th joint meeting of the European
  software engineering conference and the ACM SIGSOFT symposium on The
  foundations of software engineering}, pages 91--100. ACM, 2009.

\end{thebibliography}



\end{document}